\documentclass[prx,aps, unsortedaddress, superscriptaddress]{revtex4-2}
\pdfoutput=1
\usepackage[english]{babel}
\usepackage[utf8]{inputenc}
\usepackage{amsmath,amssymb,graphicx,hyperref}
\usepackage{centernot}

\newcommand{\mathd}{\mathrm{d}}
\newcommand{\nobracket}{}
\newcommand{\nocomma}{}
\newcommand{\textdots}{...}
\newcommand{\tmdetailed}[2]{\trivlist{\item[$\circ$]\mbox{}#2}}
\newcommand{\tmem}[1]{{\em #1\/}}
\newcommand{\tmmisc}[1]{\thanks{\textit{Misc:} #1}}
\newcommand{\tmop}[1]{\ensuremath{\operatorname{#1}}}
\newcommand{\tmrsub}[1]{\ensuremath{_{\textrm{#1}}}}
\newcommand{\tmtextbf}[1]{\text{{\bfseries{#1}}}}
\newcommand{\tmtextit}[1]{\text{{\itshape{#1}}}}
\newcommand{\tmverbatim}[1]{\text{{\ttfamily{#1}}}}
\newcommand{\nonconverted}[1]{\mbox{}}

\usepackage[OT2,T1]{fontenc}
\DeclareSymbolFont{cyrletters}{OT2}{wncyr}{m}{n}
\DeclareMathSymbol{\Sha}{\mathalpha}{cyrletters}{"58}

\begin{document}

\title{Absence of a dissipative quantum phase{\linebreak} transition in
Josephson junctions: Theory}

\author{Carles Altimiras}
\affiliation{Université \ Paris-Saclay, \ CEA, \ CNRS, \ SPEC\\
91191 \ Gif-sur-Yvette \ Cedex, France}

\author{Daniel Esteve}
\affiliation{Université \ Paris-Saclay, \ CEA, \ CNRS, \ SPEC\\
91191 \ Gif-sur-Yvette \ Cedex, France}

\author{Çağlar Girit}
\affiliation{Université \ Paris-Saclay, \ CEA, \ CNRS, \ SPEC\\
91191 \ Gif-sur-Yvette \ Cedex, France}

\author{Hélène le Sueur}
\affiliation{Université \ Paris-Saclay, \ CEA, \ CNRS, \ SPEC\\
91191 \ Gif-sur-Yvette \ Cedex, France}

\author{Philippe Joyez}
\tmmisc{philippe.joyez@cea.fr}
\affiliation{Université \ Paris-Saclay, \ CEA, \ CNRS, \ SPEC\\
91191 \ Gif-sur-Yvette \ Cedex, France}

\date{Date: January 15, 2025}

\begin{abstract}
  We investigate the resistively shunted Josephson junction (RSJ) at
  equilibrium, using linear response, an exact path integral technique and
  symmetry considerations. All three approaches independently lead to conclude
  that the superconducting{\hspace{0pt}}-{\hspace{0pt}}insulating quantum
  phase transition long believed to occur in the RSJ, cannot exist. For all
  parameters, we find that shunting a junction makes it more superconducting.
  We reveal that the UV cutoff of the resistor plays an unforeseen key role in
  these systems, and show that the erroneous prediction of an insulating state
  resulted in part from assuming it would not. We also explain why the RSJ
  physics differs from that of 1D quantum impurity problems. Our results fully
  support and confirm the experimental invalidation of this quantum phase
  transition by Murani {\tmem{et al}}. in 2020. \ 
\end{abstract}

{\maketitle}

\section{Introduction}

In the early 1980s Caldeira and Leggett {\cite{caldeira_quantum_1983}}
introduced a Hamiltonian allowing a rigorous quantum-mechanical description of
arbitrary linear circuits connected to a Josephson junction (JJ). Using this
Hamiltonian, they predicted quantitatively how dissipation reduces quantum
tunneling of the junction's phase --a macroscopic electrical variable-- and it
was precisely confirmed experimentally a few years later
{\cite{clarke_quantum_1988}}.

Shortly after Caldeira and Leggett introduced their modeling of dissipative
systems, Schmid {\cite{schmid_diffusion_1983}} predicted that a dissipative
quantum phase transition (QPT) should occur for a quantum particle in a 1D
periodic potential submitted to friction : Above a well-defined threshold in
the friction strength, independent of the potential depth and particle mass,
the particle localizes in one well of the potential, while below this
threshold it is delocalized, in apparent continuity with the Bloch states that
exist in absence of friction.

At the end of his Letter {\cite{schmid_diffusion_1983}}, Schmid briefly
mentions a resistively shunted Josephson junction (RSJ) is analogous to the
system he considers and suggests one could use it as a test bed to observe his
predicted {\tmem{localization}} effect. In this analogy, the phase of the
junction plays the role of the particle's position, the friction strength
scales as $R^{- 1}$, the inverse of the shunt resistance, and Schmid's analogy
implies the junction's phase should be localized only when the shunt
resistance $R$ is smaller than $R_Q = h / 4 e^2 \simeq 6.5 \text{k} \Omega$
and delocalized when $R > R_Q$, irrespective of the junction's characteristics
(size, transparency, material{\textdots}). The standard interpretation of this
localization\textbar delocalization dissipative QPT is that, at $T = 0$, the
JJ should be superconducting for resistances $R < R_Q$ and {\tmem{insulating}}
for $R > R_Q$. Even though this predicted insulating phase strangely conflicts
with the perturbative limit $R \rightarrow \infty$ and the classical
understanding of JJs (see Appendix \ref{history}), theoretical papers that
examined the subject using many different techniques have, to the best of our
knowledge, all essentially confirmed this interpretation
{\cite{bulgadaev_phase_1984,fisher_quantum_1985,guinea_diffusion_1985,aslangul_quantum_1985,zaikin_dynamics_1987,schon_quantum_1990,ingold_effect_1999,herrero_superconductor-insulator_2002,kimura_temperature_2004,kohler_quantum_2004,werner_efficient_2005,lukyanov_resistively_2007}}
and the phenomenon was linked with quantum impurity problems
{\cite{kane_transport_1992,torre_viewpoint_2018}}.

In 2020 Murani \tmtextit{et al.} {\cite{murani_absence_2020}} (including most
of the present authors) used state-of-the-art experimental techniques to
investigate SQUIDs, {\tmem{i.e.}} flux-tunable JJs, shunted with resistances
$R \geqslant 1.2 R_Q$. They observed a dc magnetic flux modulated the linear
response (implying the SQUID loop hosted a dc supercurrent, hence not being
insulating), but saw no trace of $T$-power-law dependence of this linear
response at low temperatures, the expected hallmark of a quantum critical
behavior {\cite{vojta_quantum_2003}}. Based on these experimental
observations, Murani \tmtextit{et al.} concluded to the absence of the
insulating state predicted by the standard interpretation of Schmid's analogy
for Josephson junctions. Subsequently, a number of papers
{\cite{morel_double-periodic_2021,masuki_absence_2022,sepulcre_comment_2022,houzet_microwave_2023,kuzmin_observation_2023,daviet_nature_2023,giacomelli_emergent_2023,paris_resilience_2024-1}}
explicitly reaffirmed the existence of Schmid's ``insulating state'' in
Josephson junctions, at least in some parameter domain. Thus, the scientific
community has not yet attained a consensus regarding the presence or absence
of Schmid's QPT in JJs. This underscores the current lack of a comprehensive
theoretical understanding of the RSJ.

In this work, we bring full theoretical support to the conclusion of Murani
\tmtextit{et al.} that RSJs are always superconducting in their ground state.
To do so, we model the RSJ using the standard the Caldeira-Leggett
Hamiltonian. By first applying Kubo's linear response theory, we indeed simply
prove that, at equilibrium, the junction is superconducting for all
parameters. Then, independently, we use an exact method based on the path
integral formalism, from which we can predict the junction's state and
transport properties. Already at the qualitative level of the path integral
equations, we can rule out the existence of the predicted QPT, and we relate
this to a ground state degeneracy of that model Hamiltonian that was not
previously appreciated. Our numerical results show that, for all parameters
tested, a resistively shunted Josephson junction is always more
superconducting than the same unshunted junction. By highlighting differences
between works predicting the QPT on one side and our present work and
experiments on the other side, we elucidate how they came to predict a quantum
phase transition.

\section{Formulation of the problem}

\begin{figure}[h]
  \resizebox{8cm}{!}{\includegraphics{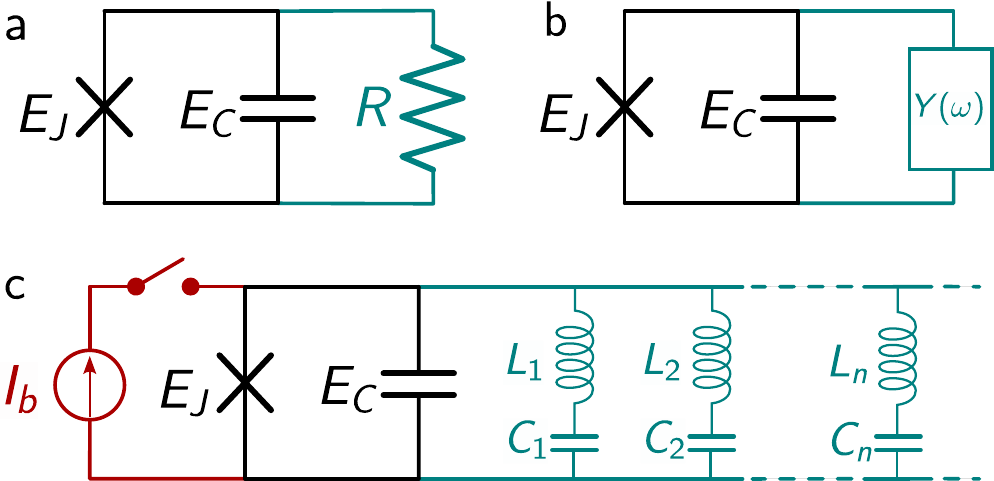}}
  \caption{\label{schem} Various schematics of a RSJ. All panels : in black,
  the junction (cross symbol) and its intrinsic capacitance, forming together
  a Cooper pair box. In blue, a dissipative element shunting the junction.
  \tmtextbf{a:} the dissipative element is generally depicted as a resistor
  \tmtextbf{b}: A resistor always has a high frequency cutoff.
  Correspondingly, the complete frequency-dependent behavior is modeled here
  by an admittance $Y (\omega)$. \tmtextbf{c}: In the Caldeira-Leggett
  Hamiltonian, the admittance is decomposed as shown : an infinite ensemble of
  $L C$ resonators at all frequencies. For figuring the linear response of the
  RSJ, we may connect a current source as depicted in red.}
\end{figure}

We model the effect of dissipation on a Josephson junction in the same way as
Caldeira and Leggett (CL) {\cite{caldeira_quantum_1983}}, with a bath of LC
harmonic oscillators providing a linear viscous damping force proportional to
the voltage across the junction ({\tmem{i.e.}} the time derivative of the
junction's phase), independently of the value of the phase (See Fig.
\ref{schem}). The corresponding RSJ Hamiltonian is
\begin{equation}
  \label{HCL} H_0 = E_C N^2 - E_J \cos \varphi + \sum_n 4 e^2  \frac{N_n^2}{2
  C_n} + \varphi_0^2  \frac{(\varphi_n - \varphi)^2}{2 L_n},
\end{equation}
where $\varphi_0 = \hbar / 2 e$ is the reduced flux quantum, $\varphi$ (resp.
$N$) denotes the junction's phase (resp. number of Cooper pairs on the
capacitor) which verify $[\varphi, N] = i$, and the $\varphi_n$ (resp. $N_n$)
denote the phase (resp. dimensionless charge) of the bath harmonic
oscillators.

The voltage operator across the parallel elements is $V = \frac{2 e}{C} N$ and
the operator for the current in the junction is $I_J = I_0 \sin \varphi$, with
$I_0 = E_J / \varphi_0$. The Hamiltonian \eqref{HCL} yields the equations of
motion
\begin{eqnarray}
  V & = & \varphi_0 \dot{\varphi}  \label{eomv}\\
  I_J & = & - 2 e \left( \dot{N} + \sum_n \dot{N}_n \right),  \label{eomi}\\
  \dot{I}_J & = & \frac{E_J}{2 \varphi_0^2} (\cos \varphi V + V \cos \varphi)
  .  \label{eomidot}
\end{eqnarray}
The first line is just the Josephson relation, the second line expresses
current conservation, and the third line evidences an inductive character of
the junction ($\dot{I}_J \propto V$), at the operator level.

\subsection{Kubo linear response of the RSJ}\label{KuboTh}

We first use Kubo's linear response theory
{\cite{kubo_statistical-mechanical_1957-1}} to evaluate the transport
properties of the system at equilibrium. Thus, we consider perturbing $H_0$ by
adding a small current-bias term $- \varphi_0 \varphi I_b (t)$, and seek the
response expectation value $\mathcal{V} (t)$ (respectively, $\mathcal{I}_J
(t)$) of the voltage across the RSJ (resp., the current flowing in the
junction) at first order in the perturbation. A straightforward application of
Ref. {\cite{kubo_statistical-mechanical_1957-1}} gives
\begin{eqnarray*}
  \mathcal{V} (t) = \int_{- \infty}^{+ \infty} z (t - t') I_b (t') d t' & ; &
  \mathcal{I}_J (t) = \int_{- \infty}^{+ \infty} \chi (t - t') I_b (t') d t'
\end{eqnarray*}
with the impulse response functions
\begin{eqnarray}
  z (t) & = & - \frac{i}{\hbar} \theta (t) \langle [\varphi_0 \varphi, V (t)]
  \rangle = - i \frac{R_Q}{2 \pi} \theta (t) \langle [\varphi, \dot{\varphi}
  (t)] \rangle,  \label{zt}\\
  \chi (t) & = & - \frac{i}{\hbar} \theta (t) \langle [\varphi_0 \varphi, I_J
  (t)] \rangle = i \theta (t) \langle [\varphi, \dot{N} (t)] \rangle .
  \nonumber
\end{eqnarray}
In the above expressions, $\mathcal{V} (t) = \tmop{Tr} (\Delta \rho (t) V)$
and $\mathcal{I}_J (t) = \tmop{Tr} (\Delta \rho (t) I_J)$, with $\Delta \rho
(t)$ the change of the density matrix at lowest order in $I_b (t)$, $\langle
\bullet \rangle = \tmop{Tr} (\rho_{\tmop{eq}} \bullet)$ with
$\rho_{\tmop{eq}}$ the unperturbed density matrix and $\bullet$ a generic
placeholder, $\theta (t)$ is the Heaviside step, and operators (except $\Delta
\rho (t)$) are evolved in the interaction picture $\bullet (t) = e^{i H_0 t /
\hbar} \bullet e^{- i H_0 t / \hbar}$. For obtaining the final writing of $z
(t)$ and $\chi (t)$, we made use of the equations of motion
(\ref{eomv}-\ref{eomi}), suited in this interaction picture, and of $[\varphi,
N_n] = 0$.

In the following, we specifically consider a current step $I_b (t) = \theta
(t) I_b$, for which we can rewrite
\begin{eqnarray*}
  \mathcal{I}_J (t > 0) & = & i I_b \int_0^t d \tau \langle [\varphi, \dot{N}
  (\tau)] \rangle\\
  & = & I_b (1 + i \langle [\varphi, N (t)] \rangle) .
\end{eqnarray*}
In the second line, we performed the integration, and used $[\varphi, N] = i$.
We now evaluate the dc value $\overline{\mathcal{I}_J}$ of the above result as
$ \overline{\bullet} = \lim_{T \rightarrow \infty} \frac{1}{T} \int_{t > 0}^{t
+ T} d \tau \bullet (\tau)$, yielding
\begin{eqnarray}
  \overline{\mathcal{I}_J} & = & I_b (1 + i \langle [\varphi, \overline{N}]
  \rangle) \nonumber\\
  & = & I_b,  \label{asymptotic}
\end{eqnarray}
where $ \overline{N}$ it thus the time-invariant part of $N (t)$, which,
consequently, commutes with $H_0$ and $\rho_{\tmop{eq}}$. In the last step, we
used $ \langle [\varphi, \overline{N}] \rangle = \tmop{Tr} \rho_{\tmop{eq}}
[\varphi, \overline{N}] = \tmop{Tr} [\overline{N}, \rho_{\tmop{eq}}] \varphi =
0$. Hence, at long times, at lowest order in $I_b$, the source's current
entirely flows through the junction. Current conservation implies that no dc
current flows through the resistor and Ohm's law then dictates that no dc
voltage develops across the elements; the dc impedance $\overline{\mathcal{V}}
/ I_b = 0$ is zero.

Still for this step current bias, from the above equations, we work out the
derivatives at $t = 0^+$ :
\begin{eqnarray*}
  \frac{d\mathcal{V}}{d t} (t = 0) & = & - \frac{i}{\hbar} I_b \langle
  [\varphi_0 \varphi, V] \rangle\\
  & = & \frac{I_b}{C}
\end{eqnarray*}
and, using \eqref{eomidot},
\begin{eqnarray*}
  \frac{d^2 \mathcal{I}_J}{d t^2} (t = 0) & = & - \frac{i}{\hbar} \varphi_0
  I_b \langle [\varphi, \dot{I}_J] \rangle\\
  & = &  \frac{E_J}{\varphi_0^2} \frac{I_b}{C} \langle \cos \varphi \rangle\\
  & = & \frac{E_J}{\varphi_0^2} \langle \cos \varphi \rangle
  \frac{d\mathcal{V}}{d t} (0)
\end{eqnarray*}
which confirms that the junction is inductive, with, at equilibrium, an
effective linear inductance
\begin{equation}
  L_{\tmop{eff}} = \varphi_0^2 / E_J \langle \cos \varphi \rangle
  \label{Leff},
\end{equation}
as already derived in Ref. {\cite{murani_absence_2020}}. The above long-time
dc superconducting response is due to this inductive behavior.

Thus, Kubo's approach yields a superconducting dc linear response for all
parameters of this Hamiltonian, and, remarkably, this result is independent of
most system details. It does not depend on the actual presence of the
dissipative bath, and applies for any lumped circuit where a generalized
inductive element (\tmtextit{i.e.} with a phase-dependent energy) is in
parallel with a capacitor (even an intrinsic or a parasitic one),
\tmtextit{e.g.} a Cooper pair box, a weak link, a parallel RLC
circuit,{\textdots} As well, the dc result \eqref{asymptotic} depends only on
operator relations, not on the system's equilibrium {\tmem{state}}. Thus, were
a QPT to occur in that system, it would not manifest as a
superconducting-to-insulating transition. Later in this article we show with
other arguments that, moreover, the localization\textbar delocalization QPT
predicted by Schmid does not occur in this model of the RSJ.

The above Kubo linear response is often considered in the frequency domain.
Taking the Fourier transform of \eqref{zt} and making use of the detailed
balance, one obtains a form of the fluctuation-dissipation theorem
\[ \tmop{Re} Z (\omega) = \frac{R_Q}{4 \pi} (1 - e^{- \beta \hbar \omega})
   \omega S_{\varphi \varphi}  (\omega) \]
where $Z (\omega) = \int z (t) e^{i \omega t} d t$ is the zero-bias linear
impedance of the RSJ as seen from the current source, and $S_{\varphi \varphi}
(\omega)$ is the spectral density of phase fluctuations, \tmtextit{i.e.} the
Fourier transform of the correlator $\langle \varphi (t) \varphi \rangle$.
Schmid {\cite{schmid_diffusion_1983}}, Schön and Zaikin
{\cite{schon_quantum_1990}}, and many other authors, determine the insulating
or superconducting character of the junction by evaluating the so-called
{\tmem{dc mobility}} $\lim_{\omega \rightarrow 0} (\omega S_{\varphi \varphi} 
(\omega))$. This quantity is indeed proportional to $\tmop{Re} Z (\omega = 0)
= \overline{\mathcal{V}} / I_b$, provided one ensures $\hbar \omega \gg k T$
while taking the limit. Clearly, the non-zero dc mobility for $R > R_Q$
predicted by Schmid {\cite{schmid_diffusion_1983}} and confirmed countless
times in the literature, is inconsistent with the null result obtained above
in the time-domain. Later, we explain what could lead to finding a transition
to a non-zero value of the mobility.

\subsection{Detailed description of the RSJ}

In order to actually provide dissipation, the bath oscillators in \eqref{HCL}
are in infinite number, forming a continuum in the frequency domain,
characterized by the spectral density of modes
\[ J (\omega) = \frac{\pi}{2}  \sum_{n = 1}^N \omega_n^2 Y_n \delta (\omega -
   \omega_n) = \omega \text{Re} Y (\omega), \]
where $\omega_n = 1 / \sqrt{L_n C_n}$ is the $n^{\tmop{th}}$ mode angular
frequency, $Y_n = \sqrt{C_n / L_n}$ its admittance, and $Y (\omega)$ the
admittance formed by the continuum. Although this model and the numerical
technique we employ below can handle any form of the admittance, we will focus
here on the so-called Ohmic case where Re$Y (\omega = 0) = 1 / R$, with $R$
the dc shunting resistance, such that $J (\omega)$ is linear in frequency at
low frequency. For fundamental reasons, any concrete dissipative bath has a UV
cutoff frequency {\cite{caldeira_quantum_1983}}. Here, we assume that Re$Y
(\omega)$ has a Lorentzian shape
\begin{equation}
  \text{Re} Y (\omega) = \frac{R^{- 1}}{1 + (\omega / \omega_c)^2} \label{ReY}
\end{equation}
which would correspond to a $LR$ series circuit, with $\omega_c = R / L$. In a
practical implementation of a metallic resistor, the inductance $L$ would be
at least the geometrical inductance of the device. Quantitative predictions on
the system will depend on the precise shape of the cutoff, but, when only
qualitative understanding is sought, it may be simpler to reason with other
cutoff shapes (e.g. abrupt or exponential). In any case, our modeling enables
considering the theoretical $\omega_c \rightarrow \infty$ limit.

The quadratic forms where the junction's phase appears in the last term of
(\ref{HCL}) can be expanded, giving
\[ H_0 = H_{\tmop{CPB}} + H_{\tmop{bath}} + H_{\tmop{coupling}} +
   H_{\tmop{CT}} \]
with the different parts corresponding, respectively, to a bare Cooper pair
box (CPB)
\begin{equation}
  H_{\tmop{CPB}} = E_C N^2 - E_J \cos \varphi, \label{HCPB}
\end{equation}
the uncoupled bath of harmonic oscillators
\[ H_{\tmop{bath}} = \sum_n \frac{(2 eN_n)^2}{2 C_n} + \frac{(\varphi_0
   \varphi_n)^2}{2 L_n} = \sum_n \hbar \omega_n  \left( a_n^+ a_n +
   \frac{1}{2} \right), \]
the coupling term
\[ H_{\tmop{coupling}} = - \varphi_0 \varphi \times \left( \sum_n \varphi_0 
   \frac{\varphi_n}{L_n} \right) = - \varphi_0 \varphi \times I_Y \]
where the junction phase $\varphi$ couples to the current $I_Y$ flowing in the
admittance $Y (\omega)$, and the so-called counter-term
\begin{eqnarray*}
  H_{\tmop{CT}} & = & (\varphi_0 \varphi)^2  \sum_n \frac{1}{2 L_n}\\
  & = & (\varphi_0 \varphi)^2  \int_0^{\infty} \frac{\mathd \omega}{\pi} 
  \text{Re} Y (\omega) = \frac{(\varphi_0 \varphi)^2}{2 L}\\
  & = & E_L \varphi^2,
\end{eqnarray*}
which appears as a parabolic inductive potential term for the junction phase
and which is essential for having the expected damped equations of motion in
the classical limit {\cite{caldeira_quantum_1983}}. Interestingly, the
counter-term transforms our CPB Hamiltonian into a fluxonium
{\cite{manucharyan_fluxonium_2009-2}} Hamiltonian at zero external flux
\[ H_{\tmop{CPB}} + H_{\tmop{CT}} = H_{\tmop{Fl}} . \]
At this point we highlight that the coupling term scales as $1 / R$, making it
perturbative in the large $R$ limit. Note that if one considers a fixed cutoff
frequency, the counter-term inductive energy $E_L$ also vanishes as $1 / R$
(since $1 / L = \omega_c / R$). Thus, in this Caldeira-Leggett model, a very
large shunt resistor appears as a perturbation to the CPB, in agreement with
the intuitive expectation that when $R$ increases to infinity no current can
flow into it, so that dissipation disappears and one can just remove the
resistance from the circuit. In the case of a purely inductive shunt with $L
\rightarrow \infty$, one also recovers the physics of a CPB
{\cite{koch_charging_2009}} (but we will not appeal to this result in the
following).

\section{Equilibrium reduced density matrix from path integrals}\label{RDM-PI}

The equilibrium reduced density matrix (RDM) of the CPB (i.e. the junction and
its capacitor) at temperature $T$ is obtained as
\begin{equation}
  \rho_{\beta} = \frac{1}{Z} \textrm{Tr\tmrsub{b}} \hspace{0.27em} e^{- \beta
  H}, \label{eq:rho_reduced}
\end{equation}
where $\beta = (k_B T)^{- 1}$ is the inverse temperature, $Z = \mathrm{Tr}
[\exp (- \beta H)]$ is the partition function of the entire system, and
$\tmop{Tr}_b$ corresponds to tracing out the bath oscillators. For the linear
coupling term and the harmonic bath we have, this tracing out can be performed
exactly, yielding the matrix elements of the RDM in phase representation as a
path integral in imaginary time
{\cite{caldeira_quantum_1983,feynman_theory_1963,grabert_quantum_1988,weiss_quantum_2012}}
\begin{equation}
  \rho_{\beta}  [\phi, \phi'] = \frac{1}{Z}  \int \mathcal{D} \varphi \exp
  \left[ - \frac{1}{\hbar} (S_{\tmop{Fl}}^E [\varphi] + \Phi [\varphi])
  \right], \label{RDMaction}
\end{equation}
where the functional integral is over all imaginary time paths $\varphi
(\tau)$ having the boundaries $\varphi (0) = \phi$ and $\varphi (\hbar \beta)
= \phi'$. In this expression, the terms in the exponential respectively denote
the Euclidean action of the fluxonium
\begin{equation}
  S_{\tmop{Fl}}^E [\varphi] = \int_0^{\hbar \beta} d \tau
  \mathcal{L}_{\tmop{Fl}} [\varphi], \label{SFl}
\end{equation}
with $\mathcal{L}_{\tmop{Fl}} [\varphi] = \frac{\hbar^2}{4 E_C} 
\dot{\varphi}^2 - E_J \cos \varphi + E_L \varphi^2$, the Lagrangian of the
fluxonium and
\begin{equation}
  \Phi [\varphi] = - \frac{1}{2}  \int_0^{\hbar \beta} d \tau \int_0^{\hbar
  \beta} d \tau' \varphi (\tau) K (\tau - \tau') \varphi (\tau'), \label{IF}
\end{equation}
the Feynman-Vernon influence functional {\cite{feynman_theory_1963}}, with the
kernel
\begin{equation}
  K (\tau) = \frac{R_Q}{2 \pi} \mathcal{C}_{I \nocomma I}  (- i \tau)
  \label{K}
\end{equation}
where $\mathcal{C}_{I \nocomma I}$ is the equilibrium autocorrelation function
of the current in the admittance (shunted at its ends). For $t \in
\mathbb{R}$, $\mathcal{C}_{I \nocomma I} (t)$ is obtained using the quantum
fluctuation-dissipation theorem and the Wiener-Khinchin theorem
\begin{equation}
  \mathcal{C}_{I \nocomma I} (t) = 2 \int_{- \infty}^{\infty} \hbar \omega
  \text{Re} Y (\omega) \frac{e^{- it \omega}}{(1 - e^{- \beta \hbar \omega})} 
  \hspace{0.17em} \frac{d \omega}{2 \pi} \label{SII}
\end{equation}
which shows that without a UV cutoff in Re$Y (\omega)$, $\mathcal{C}_{I
\nocomma I}  (t)$ would be divergent for all $t \in \mathbb{R}$ and hence
nonphysical. In Eq. \eqref{K} this expression is simply prolonged to complex
times, yielding
\begin{equation}
  K (\tau) = \frac{R_Q}{2 \pi}  \int_0^{+ \infty} \hbar \omega \text{Re} Y
  (\omega) \frac{2 \cosh \left[ \left( \frac{\beta \hbar}{2} - \tau \right)
  \omega \right]}{\sinh \frac{\beta \hbar \omega}{2}}  \frac{d \omega}{2 \pi}
  \label{Kernel},
\end{equation}
and one can check that $\int_0^{\beta \hbar} d \tau K (\tau) = 2 E_L$. In
Appendix \ref{noise_gen}, we provide analytical expressions for $K (\tau)$,
for the Lorentzian Re$Y (\omega)$ we consider. In Appendix \ref{actions}, we
show that our action is consistent with that used by Schmid and other authors,
although we have a more general kernel in the influence functional.

\subsection{Hubbard-Stratonovich transformation}

For evaluating the path integral \eqref{RDMaction}, we then rewrite the
influence functional by means of a Hubbard-Stratonovich
{\cite{hubbard_calculation_1959,stratonovich_method_1957}} transformation. In
this process, one introduces an auxiliary random scalar field $\xi (\tau)$
having Gaussian fluctuations verifying
\begin{equation}
  \langle \xi (\tau) \xi (\tau') \rangle =\mathcal{C}_{I \nocomma I}  (- i
  (\tau - \tau')) \label{ksiksi},
\end{equation}
such that the double integral in Eq. (\ref{IF}) involving $\varphi$ at two
different imaginary times can be replaced by a single integral involving
$\varphi$ at only one time, averaged over all possible realizations of $\xi$
{\cite{stockburger_exact_2002,moix_equilibrium-reduced_2012,mccaul_partition-free_2017}}.
Upon this exact transformation, Eq. \eqref{RDMaction} becomes :
\[ \rho_{\beta}  [\phi, \phi'] = \frac{1}{Z}  \int \mathcal{D} \xi
   \hspace{0.27em} W [\xi]  \int \mathcal{D} \varphi \exp \left[ -
   S_{\tmop{Fl}}^E [\varphi] - \frac{1}{\hbar}  \int_0^{\hbar \beta} d \tau
   \xi (\tau) \varphi_0 \varphi \right], \]
with a Gaussian weight functional $W [\xi]$ ensuring Eq.\eqref{ksiksi}. In the
last expression, the terms in the exponential can be seen as the Euclidean
action of a fictitious system made of a fluxonium coupled to a given
realization of a random ``current noise'' $\xi (\tau)$ generated by the bath,
so that Eq. \eqref{RDMaction} is now reformulated as
\begin{equation}
  \rho_{\beta}  [\phi, \phi'] = \frac{1}{Z}  \int \mathcal{D} \xi
  \hspace{0.27em} W [\xi]  \int \mathcal{D} \varphi \exp \left[ -
  \frac{1}{\hbar} S_{\tmop{Fict}}^E [\varphi, \xi] \right], \label{PI2}
\end{equation}
with
\begin{equation}
  \label{Sfict} S_{\tmop{Fict}}^E  [\varphi, \xi] = \int_0^{\hbar \beta} d
  \tau \left( \frac{\hbar^2}{4 E_C}  \dot{\varphi}^2 - E_J \cos \varphi + E_L
  \varphi^2 + \xi \varphi_0 \varphi \right) .
\end{equation}

\subsubsection{Invalidation of Schmid's QPT}

At this point, one can realize that valid states for these equations all have
a finite extent in $\varphi$. Indeed, in the action of Eq. \eqref{Sfict}, when
$0 < E_L = \hbar \omega_c R_Q / 4 \pi R$, the counter-term $E_L \varphi^2$
acts as a confining potential since it dominates other terms at large $|
\varphi |$ (the random noise $\xi$ is $\varphi$-independent, and
Gaussian-distributed with a finite variance $\mathcal{C}_0$ for its mean value
given by \eqref{kernelFourier}). Hence, as long as $\omega_c > 0$ and $R <
\infty$, the ground state of the action (\ref{PI2}-\ref{Sfict}) is localized
for all parameters. Given the link between Schmid's action and ours (see
Appendix \ref{actions}), the localized ground state yielded by our equations
is also a valid ground state for his action, for any $R < \infty$. This rules
out the ground state localization\textbar delocalization transition Schmid
predicts at $R = R_Q$.

The always-localized states we obtain may seem to break the discrete
translational symmetry present in the Caldeira-Leggett Hamiltonian \eqref{HCL}
and to be in conflict with the intuitive understanding of Schmid's prediction
presented in the Introduction that, at weak damping, the states of the system
should resemble the Bloch states that exist in absence of damping. In Appendix
\ref{symmetries}, we show that there is actually no problem there: the
translational symmetry of the Hamiltonian makes the states infinitely
degenerate in that system, such that, for $R < \infty$, one can exhibit
infinitely many ground states, either localized (like ours) or non-dispersing
Bloch-like delocalized with respect to the junction's phase. However, since
the junction's phase is not measurable, all these ground states are
indiscernible, which makes discussing about a localization\textbar
delocalization QPT futile.

In the following we show that our path integral approach enables for the
first time to make quantitative numerical predictions for the RSJ. We
illustrate this for various parameters, with some qualitative understanding of
the observed variations. In the process, we can explain why previous authors
came to predict a QPT.

\subsection{Stochastic Liouville equations}\label{SLEsec}

For any given realization of $\xi (\tau)$ in Eq. \eqref{PI2}, the integral of
the action of the fictitious system over all $\varphi$ paths can be seen as an
element $\rho_{\xi}  [\phi, \phi']$ of a (non-normalized) RDM obeying the
imaginary-time stochastic Liouville equation
\begin{equation}
  - \hbar \frac{\partial}{\partial \tau} \rho_{\xi} = (H_{\tmop{Fl}} + \xi
  (\tau) \varphi_0 \varphi) \rho_{\xi} \label{SLE}
\end{equation}
of the fictitious fluxonium coupled to the noise source $\xi (\tau)$, so that
\eqref{PI2} reads
\[ \rho_{\beta}  [\phi, \phi'] = \frac{1}{Z}  \int \mathcal{D} \xi
   \hspace{0.27em} W [\xi] \rho_{\xi}  [\phi, \phi'] . \]
The later equation translates into a path integral equation for the RDM
operators, independently of any choice of basis
\begin{equation}
  \rho_{\beta} = \frac{1}{Z}  \int \mathcal{D} \xi \hspace{0.27em} W [\xi]
  \rho_{\xi} \label{StoPI} .
\end{equation}
For obtaining the physical equilibrium RDM of the CPB one then needs to
perform the remaining path integral over $\xi$ in Eq. \eqref{StoPI}. This can
be done using the following scheme. For a given realization of $\xi (\tau)$,
one starts with $\rho_{\xi}  (\tau = 0)$ an equipartitioned diagonal matrix
(corresponding to an infinite temperature state of the fictitious fluxonium,
appropriate for $\tau = 0$) and integrates \eqref{SLE} up to $\rho_{\xi} 
(\tau = \hbar \beta)$. This yields a non-normalized RDM matrix with no
particular physical meaning. Repeating this numerical integration for a
suitable number of drawings of the random noise obeying Eq. \eqref{ksiksi}
amounts to sampling $W [\xi]$, and the normalized average of the different
$\rho_{\xi}  (\hbar \beta)$ is expected to converge to the physical
equilibrium RDM
\[ \frac{\sum \rho_{\xi}  (\hbar \beta)}{\tmop{Tr} \sum \rho_{\xi}  (\hbar
   \beta)} \rightarrow \rho_{\beta} . \]
We stress that if this stochastic averaging converges properly, the resulting
density matrix is exact; it takes into account the interaction of the system
and the bath to all orders with no approximation. Let us also note that the
above path integral method can be applied to any open system at equilibrium
where position-like degrees of freedom are linearly coupled to a linear bath.
It can be extended to cases where the system-bath coupling is a non-linear
function of the system's coordinates {\cite{mccaul_partition-free_2017}}. It
can even be extended to real-time out-of-equilibrium dynamics of the system
{\cite{stockburger_exact_2002,mccaul_partition-free_2017}} at the price of
introducing additional complex cross-correlated real-time stochastic
variables.

\subsubsection{Numerical implementation}\label{implementation}

For the numerical implementation of the above stochastic method, we choose as
working basis the $\mathcal{K}$ lowest eigenstates $\{| \Psi_k \rangle, 0
\leqslant k \leqslant \mathcal{K}- 1\}$ of the uncoupled fluxonium (the
expected finite extent of the ground state in $\varphi$ ensures that such
truncation is possible). For obtaining these eigenstates, we use an
intermediate discretized phase basis $\{\varphi_j = j \delta \varphi, \delta
\varphi \ll 2 \pi, j \in \mathbb{Z}$, $|j| < \varphi_{\max} / \delta
\varphi\}$, with $N^2 = - \partial^2 / \partial \varphi^2$ approximated as a
finite difference, so that the Hamiltonian is a tridiagonal matrix in this
discretized phase basis. Optimized diagonalization routines yield the first a
few hundred eigenstates of such tridiagonal matrices very fast, even when $\pm
\varphi_{\max}$ spans many wells of the cosine (low $E_L$).

Note that our working basis is {\tmem{very}} different from that of the bare
CPB which is the reference system we are interested in; this fluxonium basis
has notably a much greater density of levels {\cite{koch_charging_2009}}. At
low temperature, the most relevant energy scale for the bare CPB is its
transition energy from the ground state to the first exited state $\hbar
\omega_{01} = E_1 - E_0$ at zero offset charge (see Appendix \ref{CPBApp}),
which varies from $\hbar \omega_{01} \simeq E_C$ when $E_C \gg E_J$ to $\hbar
\omega_{01} \simeq \sqrt{E_J E_C}$ in the opposite limit $E_J \gg E_C$. This
is the ``natural'' energy scale we consider in the following, not the
transition frequencies of the fluxonium. We choose the truncation of the
working basis to encompass all the energy scales we take into account (and
$\varphi_{\max}$ in the intermediate basis is set accordingly).

Then, in the working basis, the stochastic differential Liouville equation
\eqref{SLE} is numerically integrated using discrete imaginary times steps
\{$\tau_m = m \delta \tau$\}, with $\delta \tau = \hbar \beta / M$ and $0
\leqslant m \leqslant M - 1$, and starting with $\rho (\tau = 0) =
I_{\mathcal{K}} /\mathcal{K}$, with $I_{\mathcal{K}}$ the identity matrix. The
actual approximate integration of \eqref{SLE} is performed using the symmetric
Trotter iteration scheme
\begin{equation}
  \rho_{\xi} (\tau_{m + 1}) = \exp \left( \varphi_0 \varphi \xi (\tau_m)
  \frac{\delta \tau}{2 \hbar} \right) . \exp \left( - H_{\tmop{Fl}} 
  \frac{\delta \tau}{\hbar} \right) . \exp \left( \varphi_0 \varphi \xi
  (\tau_m) \frac{\delta \tau}{2 \hbar} \right) . \rho_{\xi} (\tau_m)
  \label{trotter}
\end{equation}
that preserves the positivity of the RDM at each step
{\cite{riesch_analyzing_2019}}. In Appendix \ref{noise_gen}, we explain how we
generate the random noises $\xi (\tau_m) \delta \tau$ entering this iteration
scheme.

As explained above, after numerically integrating Eq. \eqref{SLE} for $P$
different realizations of $\xi$, we take the average RDM as
\begin{equation}
  \bar{\rho} = \frac{\sum_{p = 1}^P \rho_{\xi_p}  (\hbar \beta)}{\tmop{Tr}
  \sum_{p = 1}^P \rho_{\xi_p}  (\hbar \beta)} . \label{rhoav}
\end{equation}
In the large $P$ limit, this averaged RDM is expected to tend to the true
equilibrium RDM, which must be Hermitian and positive-semidefinite. After a
finite number of drawings, $\bar{\rho}$ is not perfectly Hermitian-symmetric,
however, it is legitimate to symmetrize it. Indeed, for the problem we
consider and in the basis we use, for a given drawing of the $\{\xi
(\tau_m)\}$ yielding $\rho_{\xi}$, drawing the reversed sequence $\{\xi
(\tau_{M - 1 - m})\}$ is equally probable and would yield the transposed of
$\rho_{\xi}$ (in our working basis, all the matrices in \eqref{trotter} are
real). Hence, for each drawing we may just add $\rho_{\xi}$ and its transposed
matrix to our stochastic average, so that it always remain (Hermitian-)
symmetric and positive-semidefinite (up to numerical accuracy). Note that even
without such symmetrization, when the average converges properly (see below),
the asymmetry of $\bar{\rho}$ reduces as $P$ increases, such that symmetrizing
or not the RDM does not perceptibly change the expectation values of the
observables we consider below.

While obtaining the RDM, we can simultaneously evaluate expectation values of
any operator $A$, as
\begin{eqnarray*}
  \langle A \rangle = \tmop{Tr} \bar{\rho} A & = & \frac{\sum w_p \tmop{Tr}
  \hat{\rho}_p A}{\sum w_p} = \frac{\sum w_p a_p}{\sum w_p}
\end{eqnarray*}
where $w_p = \tmop{Tr} \rho_{\xi_p}  (\hbar \beta)$, $\hat{\rho}_p =
\rho_{\xi_p}  (\hbar \beta) / w_p$ is the normalized RDM resulting from the
integration of Eq. \eqref{SLE} with the $p^{\tmop{th}}$ noise realization and
$a_p$=$\tmop{Tr} \hat{\rho}_p A$ the corresponding (nonphysical) expectation
value of $A$. In this expression, the trace of the $\rho_{\xi_p}  (\hbar
\beta)$ hence appear as the weight of a given noise realization in the final
estimate of any expectation value (drawings with large traces correspond to
paths with lower action in the path integral). The error bars on the estimated
expectation value are obtained from the estimator of the variance of the
weighted average using the Central Limit Theorem and the effective number of
data points $P_{\tmop{eff}} (P) = \left( \sum w_p \right)^2 / \sum w_p^2$.

At large shunt resistance values and high temperature, the \{$w_p$\} are such
that the effective number of samples $P_{\tmop{eff}} (P)$ grows fast with the
number of drawings $P$ and the weighted means converge well. However, when
reducing $R$ ({\tmem{i.e.}} increasing the coupling to the bath) at fixed
$E_C, E_J$, $\hbar \omega_c$ and $kT$, one must increase the number of time
steps needed to keep the random increments $\xi (\tau_m) d \tau$ small enough,
but after the random walk integration of Eq. \eqref{SLE} this nevertheless
translates into an increased variance of the \{$w_p$\}, and a corresponding
reduction of $P_{\tmop{eff}}$. At some point in this increase, the weighted
estimation of the expectation values becomes dominated by the few drawings
that fall in the (positive side) tail of the $w_p$ distribution. In other
words, when $R$ is much reduced, the average is dominated by very few drawings
(and possibly a single one when $P_{\tmop{eff}} \simeq 1$ and no longer grows
substantially with $P$). This indicates that, in this case, the action has a
deep and sharp minimum representing only an extremely small volume in the
phase space of the $\xi$ noise, making the method extremely inefficient. In
such case, whether or not the method can still yield reliable estimates of
observables depends on the derivatives of those observables around this
minimum. Similarly, when reducing the temperature (all other parameters kept
fixed), the number $M$ of steps in $\tau$ also needs to be scaled up,
eventually causing the same poor statistics. The $R$ and $T$ ranges where the
statistics are poor depend on the other system parameters, and notably on the
cutoff frequency. The data presented below are all in regimes where the
estimators have small error bars, away from these problematic limits.

\section{Results}\label{sec:results}

\begin{figure}[h]
  \begin{center}
    \resizebox{10cm}{!}{\includegraphics{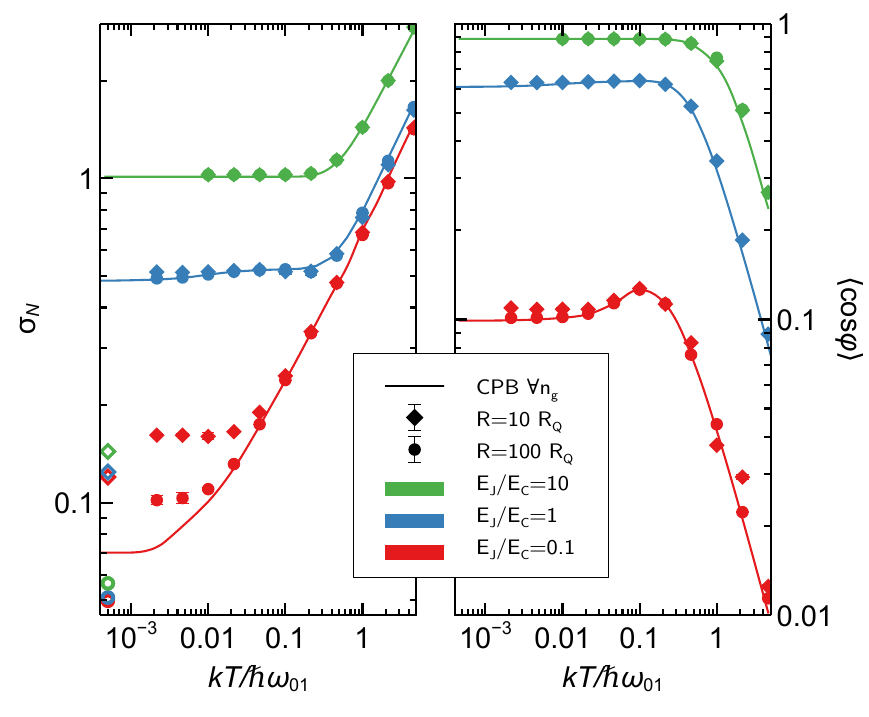}} \ 
  \end{center}
  \caption{\label{tdep}Temperature dependence of the rms charge fluctuations
  $\sigma_N$ on the capacitor (left panel) and the Josephson coherence factor
  $\langle \cos \varphi \rangle$ (right panel) for large shunt resistance
  values and different $E_J / E_C$ ratios, for $\hbar \omega_c = 0.4 \hbar
  \omega_{01}$. In both panels, the solid lines are the thermal expectation
  values for the unshunted CPB allowing any gate charge (See Appendix
  \ref{CPBApp}). For larger resistance values, the calculated expectation
  values (markers) are getting closer to the bare CPB values, as expected for
  a vanishing perturbation. Open symbols in the left panel are the zero
  temperature limits of $\sigma_{N, C || Y}$ (Eq. \eqref{sigmaNT0}) with the
  same $Y (\omega)$ (same resistance and cutoff) as the filled symbol of
  corresponding color and shape. The fact that $\sigma_N$ saturates above
  these values shows that the junction has finite supercurrent fluctuations in
  its ground state, consistent with the finite value of the Josephson
  coherence in the right panel. }
\end{figure}

In Fig. \ref{tdep} we show the expectation values of the rms charge
fluctuations $\sigma_N = \langle N^2 \rangle^{1 / 2}$ and the effective
Josephson coupling $\langle \cos \varphi \rangle$ as a function of the reduced
temperature $kT / \hbar \omega_{01}$, for different values of $E_J / E_C$, for
the RSJ at large values of $R / R_Q$. The finite values reached by these
expectation values at low temperature attest that the junction allows
(super)current flow in its ground state. Indeed, if the junction were
insulating, its effective inductance $L_{\tmop{eff}} = \varphi_0^2 / E_J
\langle \cos \varphi \rangle$ would be infinite (the Josephson coupling $E_J
\langle \cos \varphi \rangle$ vanishes) and the charge $N$ on the capacitor
would fluctuate just as in the $C || Y$ circuit, yielding
\[ \sigma_{N, C || Y} = \frac{C}{2 e} \left( \int_{- \infty}^{\infty} \hbar
   \omega \text{Re} \frac{1}{iC \omega + Y (\omega)} \coth \left( \frac{\hbar
   \omega}{2 kT} \right) \hspace{0.17em} \frac{d \omega}{2 \pi} \right)^{1 /
   2} . \]
For the Lorentzian admittance \eqref{ReY} we consider, $C || Y$ is equivalent
to a series RLC circuit, yielding the zero point fluctuations
{\cite{caldeira_quantum_1983,grabert_quantum_1984}}
\begin{equation}
  \sigma_{N, C || Y}  (T = 0) = \left( \frac{\hbar \omega_c}{4 \pi E_C} 
  \frac{\log \left( \sqrt{\alpha} + \sqrt{\alpha - 1} \right)}{\sqrt{\alpha} 
  \sqrt{\alpha - 1}} \right)^{1 / 2} \label{sigmaNT0},
\end{equation}
with $\alpha = \frac{\pi R \hbar \omega_c}{4 R_Q E_C}$. The conducting
character of the JJ is evidenced by the fact that $\sigma_N$ saturates to
values strictly larger than $\sigma_{N, C || Y}  (T = 0)$, consistently with
the finite saturation value of $\langle \cos \varphi \rangle$.

In that Fig. \ref{tdep}, we also compare our numerical results for these
observables to those obtained for the thermal averages of the bare CPB
considering all gate charge values (see Appendix \ref{CPBApp}). For these
large resistances, most of the numerical expectation values for the RSJ are
found close to that of the CPB. At low temperatures they are found slightly
above those of the CPB, but by increasing further the resistance (data not
shown) one recovers more closely the bare CPB results, as expected for a
vanishing perturbation. At large temperatures, some results for $\sigma_N$ are
slightly below the asymptote $\sqrt{kT / 2 E_C}$ (valid for both the bare CPB
and the $C || Y$ circuit), which we attribute to our basis truncation.

\begin{figure}[h]
  \begin{center}
    \resizebox{10cm}{!}{\includegraphics{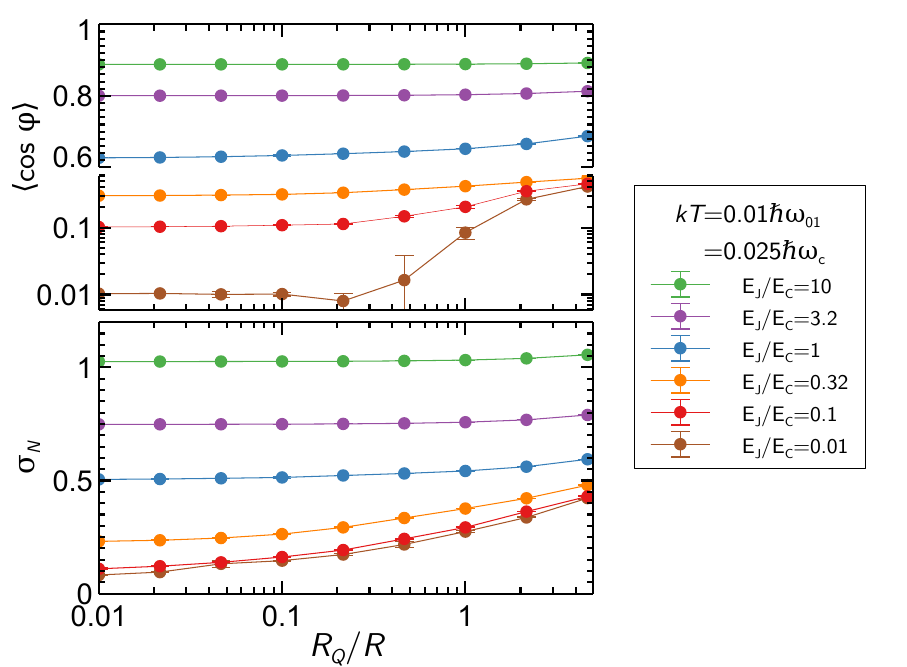}} \ 
  \end{center}
  \caption{\label{rdep}Resistance dependence of $\langle \cos \varphi \rangle$
  (top panel - note the log-lin broken vertical axis) and $\sigma_N$ (bottom
  panel) for different $E_J / E_C$ ratios at the low temperature $kT = 0.01
  \hbar \omega_{01}$ and for $\hbar \omega_c = 0.4 \hbar \omega_{01}$. One
  observes that both $\langle \cos \varphi \rangle$ and $\sigma_N$ increase
  when reducing the value of the shunt resistance and tend to saturate at the
  bare CPB value at large $R$. No change of behavior is observed around $R =
  R_Q$.}
\end{figure}

\begin{figure}[h]
  \begin{center}
    \resizebox{9cm}{!}{\includegraphics{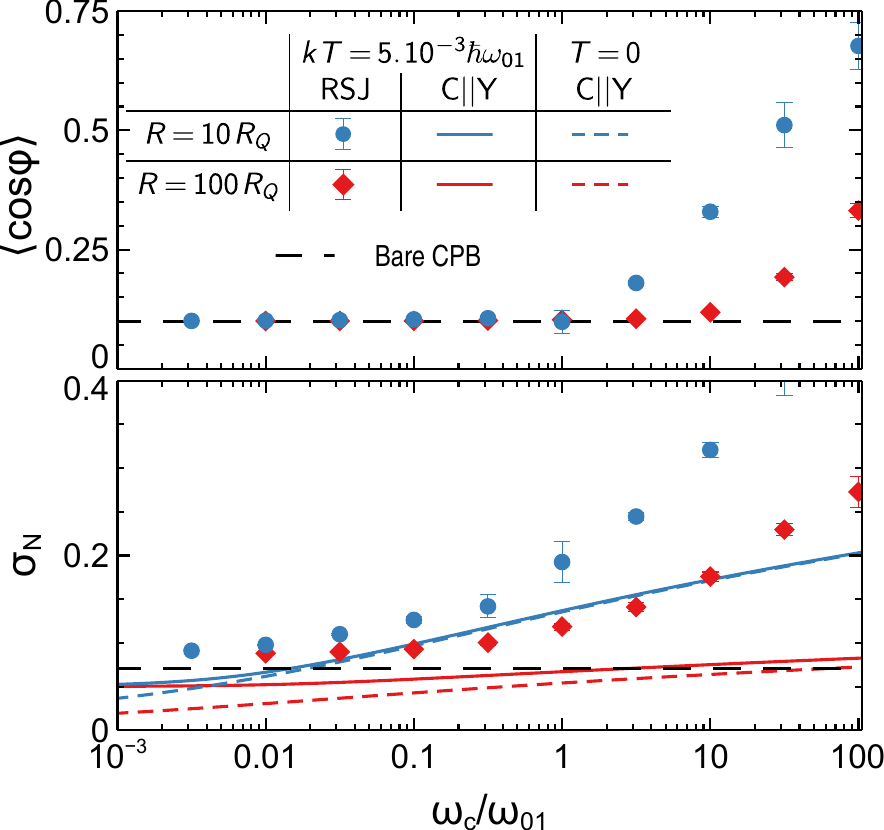}} \ 
  \end{center}
  \caption{\label{cutoffdep}Dependence of $\langle \cos \varphi \rangle$ (top
  panel) and $\sigma_N$ (bottom panel) with the cutoff frequency $\omega_c$ of
  the Ohmic bath, for an RSJ with $E_C = 10 E_J$, $R = 10 R_Q$ (blue) or $R =
  100 R_Q$ (red) at the low temperature $kT = 0.005 \hbar \omega_{01}$. The
  black dashed lines correspond to the predicted values for the ground state
  of the bare CPB. In the bottom panel the colored lines show the predicted
  charge fluctuations $\sigma_{N, C || Y}$ in absence of junction conduction,
  both at the simulated temperature (solid lines) and $T = 0$ (dashed lines).
  One observes that at low cutoff the expectation values tend to those of the
  CPB independently of $\omega_c$ while at large cutoff the expectation values
  depend on $\omega_c$, with the superconducting character of the junction
  increasing with $\omega_c$. }
\end{figure}

In Fig. \ref{rdep} we consider the $R$-dependence of the same expectation
values for different ratios $E_J / E_C$ and at the low temperature $kT = 0.01
\hbar \omega_{01}$. We observe that both $\langle \cos \varphi \rangle$ and
$\sigma_N$ smoothly increase when $R$ is reduced. In Fig. \ref{cutoffdep}, we
show that for large resistance values, large $E_C / E_J$ and at the low
temperature $kT = 0.005 \hbar \omega_{01}$, changing the cutoff frequency
$\omega_c$ of the environment admittance has a weak effect at small
$\omega_c$, while at large $\omega_c$, the expectation values of the RSJ do
depend on the actual value of the bath cutoff, the junction becoming more
superconducting as $\omega_c$ increases. Similar behavior is observed for
other $E_C / E_J$ ratios and shunt resistances values.

\begin{figure}[h]
  \resizebox{7cm}{!}{\includegraphics{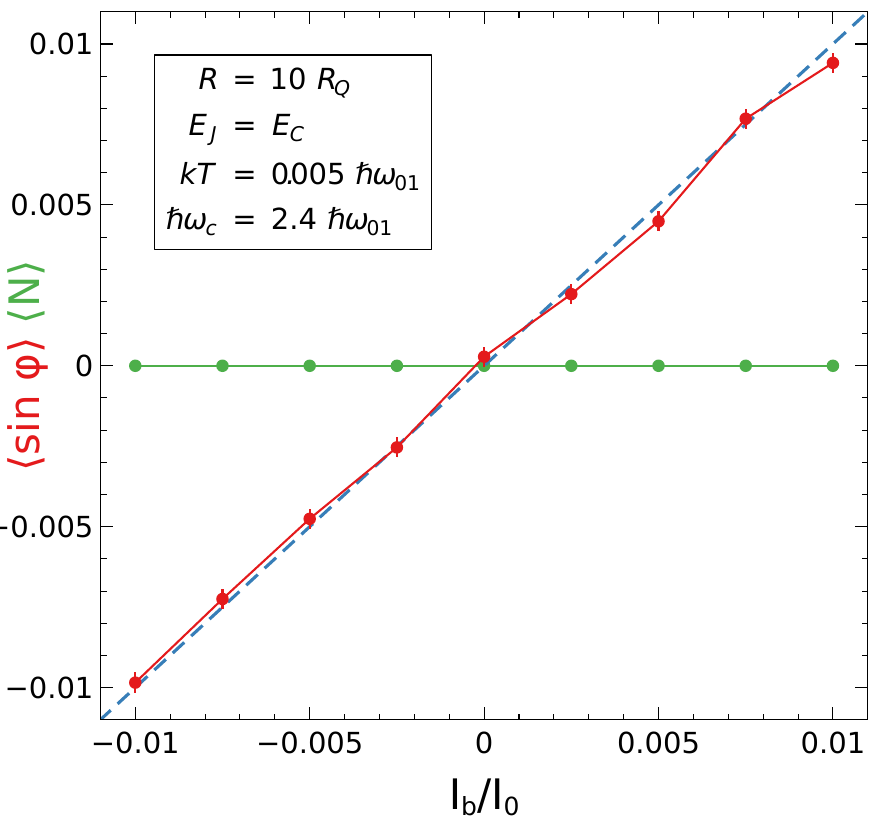}}
  \caption{\label{lin_resp}Example linear response expectations values of the
  current $I_0 \sin \varphi$ through the junction and the charge $2 e N$ on
  the capacitor (proportional to the voltage across the junction) when adding
  a small bias current $I_b \ll I_0$ to the system. With a resistance $R = 10
  R_Q$, the junction would be in Schmid's ``insulating phase'', while we
  obtain a superconducting response, in agreement with Kubo's theory (see
  \ref{KuboTh}). \ }
\end{figure}

Our method further allows to simply work out the dc linear response of the RSJ
to a current bias. Indeed, adding a dc current source $I_b$ to the system adds
a potential term $- \varphi_0 I_b \varphi$ to the Caldeira-Leggett Hamiltonian
\eqref{HCL}, and this term directly carries over to our path integrals,
shifting the minimum of the potential of the fictitious fluxonium away from
$\varphi = 0$. With this term added, for small bias current $I_b \ll I_0 = E_J
/ \varphi_0$, stochastic Liouville numerics (see Fig. \ref{lin_resp}) yield
$\langle V \rangle \propto \langle N \rangle = 0$ and $I_0 \langle \sin
\varphi \rangle = I_b$ up to numerical accuracy, corresponding to a
supercurrent flow through the junction.

\section{Discussion}\label{discussion}

The superconducting linear response shown in Fig. \ref{lin_resp} is consistent
with the result obtained from Kubo's theory in \ref{KuboTh}. It explains why
RSJ experiments {\cite{murani_absence_2020,grimm_bright_2019}} may observe
signatures of a finite dc supercurrent that saturates at low temperatures for
parameters that were previously believed to have an insulating ground state.
As said above, a small current bias slightly shifts the global minimum of the
potential of the fictitious fluxonium away from $\varphi = 0$, but the states
remain localized in phase, so that the dc voltage remains zero (this argument
is critically examined in Appendix \ref{stability}, where we also discuss what
eventually limits this superconducting behavior).

Our numerical results show that the $R \rightarrow \infty$ limit of the RSJ
smoothly recovers the well known physics of the CPB family of Josephson
qubits, as expected for a vanishing perturbation. In addition, we observe that
in the RSJ with a finite shunt resistance, at low temperatures, the effective
Josephson coupling $E_J \langle \cos \varphi \rangle$ saturates to a value
larger ($\geqslant$) than in the bare CPB and the rms charge fluctuations on
the capacitor $\sigma_N$ saturate to values larger ($\geqslant$) than in the
bare CPB, and strictly larger than in the $C || Y$ circuit, for all the
parameters we tested. This establishes that, in the Caldeira and Leggett model
with an Ohmic environment having a finite UV cutoff frequency, the shunted
Josephson junction's ground state is superconducting and actually {\tmem{more
superconducting}} than the bare CPB junction. Notably, when $E_J \ll E_C$ one
has $E_J \langle \cos \varphi \rangle \rightarrow E_J^2 / E_C$ in the CPB,
setting a lower bound to the RSJ's superconductivity at equilibrium.

Our results further show that the JJ's low-$T$ superconductivity increases at
large cutoff of the Ohmic bath. The observed trend is actually simply
explained by the counter-term localizing the phase more and more tightly
(since $E_L \propto \omega_c$), which would ultimately yield a perfectly
localized classical phase at $\varphi = 0$ (and thus $\langle \cos \varphi
\rangle = 1$) when $\omega_c = \infty$. The trend can also be equivalently
explained by the logarithmic increase with $\omega_c$ of $\sigma_{N, C || Y}$
(Eq. \eqref{sigmaNT0}), the environment-induced charge fluctuations on the
capacitor, which provides more charges states for the Josephson coupling
mechanism, driving up both $\sigma_N$ and $\langle \cos \varphi \rangle$
(eventually reaching 1). Hence, we conclude a Markovian bath would yield a
classical phase JJ with the maximal effective Josephson coupling $E_J \langle
\cos \varphi \rangle = E_J$, for all values of the resistance. Although this
result can be understood simply, it has not been realized so far, to the best
of our knowledge. Indeed, in the literature that predicted that transition, it
is widely assumed that a strictly Ohmic bath with no UV cutoff would be
appropriate for predicting the RSJ ground state (yet, not finding a fully
localized phase, for reasons explained below). This assumption that the bath's
UV cutoff would be irrelevant is most likely due to assuming that the
junction's capacitance by itself would sufficiently squash the high frequency
fluctuations in the system. However, this is not the case since charge
fluctuations on the capacitor of an RC circuit diverge at infinite cutoff (see
Eq. \eqref{sigmaNT0}), and even more so when adding a junction in parallel.

Our findings can be globally explained qualitatively by arguing that
connecting a resistor to a CPB can significantly affect the ground state of
this nonlinear oscillator only if the environment impedance $Z (\omega) = Y^{-
1} (\omega)$ is comparable to or lower than the effective impedance of the
unshunted CPB at its plasma frequency, such that it can reduce the phase
fluctuations across the junction. If furthermore the phase fluctuations of the
bare CPB are initially large ($E_C  \centernot { \ll } E_J$) the reduction of
the phase fluctuations due to the resistor leads to an increase of $\langle
\cos \varphi \rangle$, reducing the junction's effective inductance
$L_{\tmop{eff}} = \varphi_0^2 / E_J \langle \cos \varphi \rangle$, and hence
its effective impedance, which in turn bootstraps the reduction of the phase
fluctuations. Here, the method yields the exact self-consistent solution for
these environment-modified fluctuations and the corresponding effective
Josephson coupling $E_J \langle \cos \varphi \rangle$; this can be seen as
generalizing approaches restricted to Gaussian phase fluctuations (see e.g.
Ref. {\cite{joyez_self-consistent_2013}}) and that predict an effective
coupling $E_J e^{- \frac{1}{2} \langle \varphi^2 \rangle}$
{\cite{note-Gaussian-results}}. The linear impedance of the bare CPB can be
estimated using
\[ \frac{Z_{\tmop{CPB}}}{R_Q} \sim \frac{1}{2 \pi}  \sqrt{\frac{\langle
   \varphi^2 \rangle}{\langle N^2 \rangle}} \]
which would be exact for the harmonic oscillator, or as
\[ \frac{Z_{\tmop{CPB}}}{R_Q} \sim \frac{1}{R_Q} 
   \sqrt{\frac{L_{\tmop{eff}}}{C}} = \frac{1}{2 \pi}  \sqrt{\frac{2 E_C}{E_J
   \langle \cos \varphi \rangle}} \]
both of which evolve from $\frac{1}{2 \pi}  \sqrt{\frac{2 E_C}{E_J}} < 1$ when
$E_C \ll E_J$ to $\propto \frac{E_C}{E_J} \gg 1$ when $E_C \gg E_J$. This
roughly explains at which resistance value the upturn of $\langle \cos \varphi
\rangle$ occurs in Fig. \ref{rdep}. Yet, for $E_J / E_C \ll 1$, resistances
much larger than the above estimates of the CPB linear impedance already
induce a substantial change of $\sigma_N$ compared to the bare CPB, an effect
dependent on the cutoff $\omega_c$ (See Fig. \ref{cutoffdep}).

The temperature dependence of $\langle \cos \varphi \rangle$ is strikingly
non-monotonic for $E_J \ll E_C$ (see Fig. \ref{tdep}). Starting from low
temperatures, it first shows a plateau corresponding to the zero point
fluctuations, followed by an increase with a local maximum around $kT / \hbar
\omega_{01} = 0.1$, before reducing and finally vanishing at high
temperatures. We believe that this temperature dependence of $\langle \cos
\varphi \rangle$ explains the non-monotonic variation observed in the
experimental results of Ref. {\cite{murani_absence_2020}}. The measured
scattering amplitude (see Fig. 3 and Appendix E in Ref.
{\cite{murani_absence_2020}}) in that experiment is
\[ | S_{21} |^2 \propto \frac{1}{| 1 + R Y_J (\omega_{\tmop{meas}}) |^2} =
   \frac{1}{1 + \left( \frac{R}{R_Q} \frac{E_J^{\tmop{sq}} (\Phi)}{\hbar
   \omega_{\tmop{meas}}} \langle \cos \varphi \rangle \right)^2}, \]
where we assume that the junction's admittance $Y_J (\omega) \simeq 1 / i
L_{\tmop{eff}} \omega$ is dominated by its inductive behavior at the
measurement frequency $\omega_{\tmop{meas}} / 2 \pi \simeq 1 \tmop{GHz}$,
with, according to \eqref{Leff}, $L_{\tmop{eff}}^{- 1} = E_J^{\tmop{sq}}
(\Phi) \langle \cos \varphi \rangle / \varphi_0^2$ and $E_J^{\tmop{sq}} (\Phi)
= E_{J \max} | \cos (\pi \Phi / \Phi_0) |$ the Josephson coupling of the
experiment's SQUID (assumed symmetric), tuned by the applied magnetic flux
$\Phi$. A precise fitting of the experimental data with the above expression
is unrealistic because the measured scattering amplitudes were uncalibrated
{\cite{murani_absence_2020}} and the cutoff frequency of the resistive
environment was not controlled (it was not known to be a relevant parameter at
the time). Yet, plugging $\langle \cos \varphi \rangle$ for $E_J / E_C = 0.1$
as computed in Fig. \ref{tdep}, and the experimental parameters $R E_{J \max}
/ R_Q \hbar \omega_{\tmop{meas}} \sim 5 - 10$ in the above experession would
already qualitatively capture the variations of the experimental data (at
temperatures sufficiently below $T_c \sim 1.2 \text{K}$ of Al), with, notably,
the temperature of the local maximum admittance (minimum $| S_{21} |^2$) in
the experiment that reasonably corresponds to $k T = 0.1 \hbar \omega_{01}
\simeq 0.1 E_C$ for both samples. This shows our path integral approach
enables detailed quantitative comparison with future tailored experiments.

As fully expected from the qualitative argument on the existence of a
localized ground state for all parameters given in Sec. \ref{RDM-PI}, our
numerical results show no sign of Schmid's dissipative QPT in JJs. In
particular, we observe no change of behavior at or near $R = R_Q$. As well,
equilibrium observables related to transport do not follow power laws of the
temperature in the critical region of the expected QPT, which would be the
numerical signature of that QPT {\cite{vojta_quantum_2003}}.

In Appendix \ref{symmetries} we show that the symmetries of the Hamiltonian
(which, surprisingly, have not been thoroughly worked out earlier) by
themselves also preclude the existence of Schmid's QPT. We notably reveal that
the ground state of the system is infinitely degenerate, such that it does not
actually possess a well-defined symmetry that a QPT could break, unlike in the
akin spin-boson problem. This degeneracy also manifests itself in a rigorously
flat quasicharge ground band (contrary to what is frequently assumed in the
literature). The predicted insulating state was supposed to consist of the
system trapped at a local minimum of that band (\tmtextit{i.e.} the exact dual
of the dc Josephson effect); the flat band forbids it.

Having invalidated Schmid's prediction in several independent ways, we still
need to explain why the entire previous theoretical literature on that
question confirmed the prediction. As can be expected, the reason is rather
subtle; we show in App. \ref{why} that it involves two unfulfilled implicit
assumptions. The first assumption, already mentioned above, is that the bath's
UV cutoff would be irrelevant, leading to directly consider an infinite
cutoff. The second implicit assumption is that this infinite cutoff limit
would commute nicely with other limits, but we reveal it does not, yielding
results that are nonphysical for this model. The flawed results obtained that
way ultimately led to the belief that the RSJ and the 1D quantum impurity
problem are equivalent, which we dispel (see App. \ref{kane-fisher}). Our work
shows that, contrary to the past literature, it is essential to take into
account the finiteness of resistor's UV cutoff (and the ensuing non-Markovian
dynamics) for predicting the superconducting RSJ ground state correctly (with
its fluctuations), whatever the theoretical approach; it is not just a matter
of taste or convenience.

Finally, the present work provides for the first time a reliable way of
predicting the equilibrium behavior of JJs in presence of arbitrary linear
environments --even frequency-dependent ones--, provided the impedance is not
too small. In the opposite small shunting impedance regime, the approach
should be doable in the dual picture, considering the coupling of the JJ
charge with the impedance's fluctuating voltage.

\section{Conclusions}

In this work we prove that the Caldeira-Leggett Hamiltonian used to model a
resistively shunted Josephson junction has no superconducting-to-insulating
quantum phase transition, contrary to what was widely believed after Schmid's
1983 prediction. We actually provide three independent demontrations relying
respectively on Kubo's linear response theory, symmetry arguments, and an
exact path integral method.

Our path integral approach lends itself to a numerical implementation yielding
the equilibrium reduced density matrix and the expectation values of
observables of the RSJ. This provides the first workable method to predict
quantitatively the behavior of the RSJ in a wide range of parameters where
predictions were previously impossible or incorrect. The method handles
arbitrary frequency-dependent environment impedances, and, in principle, it
can be extended to dynamical situations.

Our results
\begin{itemize}
  \item fully support the conclusions of Murani \tmtextit{et al.}
  {\cite{murani_absence_2020}} that a resistive shunt with $R > R_Q$ does not
  render a Josephson junction {\tmem{insulating}}. Actually, a shunt resistor
  can only make a junction {\tmem{more superconducting}} than it would be in
  its absence.
  
  \item recover the CPB physics when the shunt resistance is made very large,
  as expected for a vanishing perturbation,
  
  \item reveal an unforeseen dependence of the junction's superconducting
  properties with the resistor's UV cutoff, which must therefore be taken into
  account for making sensible predictions for the RSJ. 
\end{itemize}
Our work reveals and clarifies several subtle issues that led to the flawed
prediction and its long trail in the literature. Namely, we
\begin{itemize}
  \item point to an issue of non-commuting limits when considering a resistor
  with infinite cutoff from the onset,
  
  \item explain why the RSJ and the 1D quantum impurity problem are not
  analogous (the latter having a QPT, indeed), contrary to what was previously
  thought,
  
  \item clarify the symmetries and degeneracies of the phase states in the
  RSJ, a theoretical question that has also long been a matter of debate in
  the community. 
\end{itemize}

\section{Acknowledgments}

We are grateful to H. Grabert, J. Stockburger, C. Ciuti, L. Giacomelli, F.
Borletto, R. Riwar, N. Roch, X. Waintal, I. Snyman, M. Houzet, O. Maillet, S.
Latil, C. Gorini and our colleagues of the Quantronics group at CEA-Saclay for
stimulating discussions, comments and helpful inputs at various stages of this
work initiated 7 years ago. This work is supported in part by ANR project
Triangle ANR-20-CE47-0011-02.

\section*{Appendices}

\appendix\section{On the interpretation of Schmid's QPT in Josephson
junctions}\label{history}

What Schmid saw as remarkable in his work {\cite{schmid_diffusion_1983}}, was
the {\tmem{localization effect}} in one well at large enough friction. Indeed,
at low friction, a delocalized particle was seen as no surprise since one
expects to recover Bloch states in the vanishing friction limit.

However, the way in which Schmid's analogy was received by physicists familiar
with the Josephson junction had a totally reversed ``surprise factor'' : The
predicted localized JJ phase was seen as run-of-the-mill since it seems just
like the classical description of the JJ which the beginner first learns. On
the contrary, the delocalized phase in the weak damping limit, which was the
vanilla situation for Schmid's original particle, became an extraordinary
situation in which the JJ would turn {\tmem{insulating}} under the action of
the resistance, even when the corresponding friction force is vanishing,
implying one must give up the notion of a perturbative effect. As well,
according to what became the standard interpretation of Schmid's analogy,
{\tmem{a JJ could only be superconducting when it experienced strong damping
of its phase}}, even if this was at odds with the well-understood classical
limit of the device in which dissipation is not {\tmem{needed}} to have a
superconducting device and which well explained abundant experimental
observations of supercurrents in large underdamped, junctions (by far the
easiest to make and measure). In addition of conflicting with perturbation
theory and the classical limit, Schmid's prediction was also at odds with the
theoretical work that had already been done on the RSJ
{\cite{ivanchenko_josephson_1969,mccumber_effect_1968,stewart_current-voltage_1968}},
and which Caldeira and Leggett had completed. Notably, Caldeira and Leggett's
work quantitatively accounts for experiments on RSJs
{\cite{clarke_quantum_1988}}, and there is no trace of Schmid's QPT in their
tunneling rate.

\section{Generation of discrete noise increments with required
correlations}\label{noise_gen}

\begin{figure}[h]
  \begin{center}
    {\hspace*{\fill}}
    
    \resizebox{7cm}{!}{\includegraphics{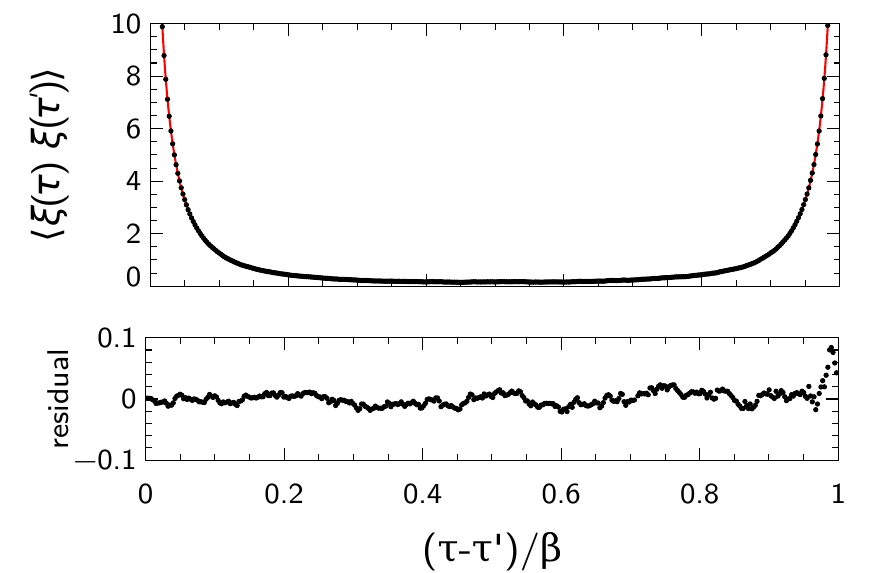}} 
  \end{center}
  \caption{\label{check_correl}Comparison of the filtered (see text)
  theoretical current noise correlator in imaginary time for a Lorentzian
  \text{Re}$Y (\omega)$ (red continuous line) and experimental correlator for
  $10^6$ drawings of a noise sequence (black dots). The bottom panel shows the
  difference between the numerical correlator and its expected value.
  Parameters are $R = 3 R_Q$, $\beta \hbar \omega_c = 30$, 401 time steps.}
\end{figure}

For the Lorentzian Re$Y (\omega)$ Eq. \eqref{ReY} we assume, the integral in
\eqref{Kernel} converges for $0 < \tau < \hbar \beta$ and admits the
analytical solutions
\begin{eqnarray}
  \mathcal{C}_{I \nocomma I}  (- i \tau) & = & \frac{\hbar}{\pi R} \omega_c^2 
  \left( \text{Re} \left[ e^{- \frac{2 i \pi \tau}{\beta \hbar}} \Phi \left(
  e^{- \frac{2 i \pi \tau}{\beta \hbar}}, 1, \frac{\beta \hbar \omega_c}{2
  \pi} + 1 \right) \right] + \frac{\pi}{\beta \hbar \omega_c} \right)
  \nonumber\\
  & = & \sum_{n = 0}^{\infty} \mathcal{C}_n  \text{cos} (\omega_n \tau) 
  \label{kernellorentz}
\end{eqnarray}
where $\Phi$ is the Lerch transcendent function, $\omega_n = n \frac{2
\pi}{\beta \hbar}$, and
\begin{equation}
  \mathcal{C}_n = \frac{\hbar}{R} \omega_c^2  \frac{(2 - \delta_{n 0})}{\beta
  \hbar \omega_c + 2 \pi n} \label{kernelFourier} .
\end{equation}
The expressions in \eqref{kernellorentz} are even and $\hbar \beta$-periodic
in $\tau$ (and, of course, symmetric about $\tau = \hbar \beta / 2$). At $\tau
\sim 0$ these expression have a mild divergence $\mathcal{C}_{I \nocomma I} 
(- i \tau) \sim R^{- 1} \omega_c^2 \log | \tau |$ (See Fig.
\ref{check_correl}). Note that other forms of Re$Y (\omega)$ with a sharper
cutoff, such as e.g. Re$Y (\omega) = \exp (- | \omega | / \omega_c) / R$, even
yield a finite $\mathcal{C}_{I \nocomma I}  (0)$, \tmtextit{i.e.} a finite
variance for $\xi (\tau )$ at all times. This should clear possible worries
associated with the mild divergence of $\mathcal{C}_{I \nocomma I}$ in the
Lorentzian case, since one expects that the overall behavior of a system
should not depend on the precise shape of such cutoff.

For satisfying Eq. \eqref{ksiksi}, the random noise $\xi (\tau_n)$ can be
naively drawn as the real numbers
\begin{equation}
  \xi (\tau_n) = \sum^{M - 1}_{m = 0} R_m  \text{cos} (\omega_m \tau_n +
  \theta_m) \label{DFT}
\end{equation}
where $\theta_0 = 0$, $R_0$ is a normally-distributed random number with zero
mean and variance $\mathcal{C}_0$, and the $\{R_{m > 0} \}$ are taken fixed as
$R_{m > 0} = \sqrt{2\mathcal{C}_m}$, with the $\{\theta_{m > 0} \}$ random and
uniformly-distributed in $[0, 2 \pi)$. Then, the $\{\xi (\tau_n)\}$ ensemble
(Eq. \eqref{DFT}) is efficiently obtained as the real part of the fast Fourier
transform (FFT) of $\{e^{i \theta_m} R_m \}$. With this construction, the
correlators are
\begin{eqnarray}
  \langle \xi (\tau_n) \xi (\tau_m) \rangle & = & \sum^{M - 1}_{j, k = 0}
  \left\langle R_k R_j  \text{cos} (\omega_k \tau_n + \theta_k) \text{cos}
  (\omega_j \tau_m + \theta_j) \right\rangle \nonumber\\
  & = & \sum^{M - 1}_{j, k = 0} \frac{1}{2} \{ \langle R_k R_j \cos (\delta
  \tau \delta \omega (jm - kn) - \theta_k + \theta_j) \rangle \nobracket
  \nonumber\\
  &  & \hspace{5em} + \langle R_k R_j \cos (\delta \tau \delta \omega (kn +
  jm) + \theta_k + \theta_j) \rangle \nobracket\} \nonumber\\
  & = & \langle R_0^2 \rangle \frac{1}{2} + \sum^{M - 1}_{j = 1} R_j^2 
  \frac{1}{2} (\cos (\delta \tau \delta \omega j (m - n))) + \langle R_0^2
  \rangle \frac{1}{2} \cos (2 \theta_0) \nonumber\\
  & = & \sum^{M - 1}_{j = 0} \mathcal{C}_j  \text{cos} (j (m - n) \delta \tau
  \delta \omega) \nonumber\\
  & = & \mathcal{C}_{I \nocomma I}  (- i (\tau_n - \tau_m)) -
  \sum^{\infty}_{j = M} \mathcal{C}_j  \text{cos} (j (m - n) \delta \tau
  \delta \omega)  \label{delta}
\end{eqnarray}
which almost fits the requirement (Eq. \eqref{ksiksi}).

The first problem with this naive algorithm is the logarithmic divergence of
$\mathcal{C}_{I \nocomma I}  (- i \tau)$ at $\tau = 0$, for the Lorentzian
Re$Y (\omega)$ we consider. This is solved by taking, instead of
$\mathcal{C}_{I \nocomma I}  (- i (\tau_n))$, the averaged
$\overline{\mathcal{C}_{I \nocomma I}} (- i (\tau_n)) = \delta \tau^{- 1} 
\int_{\tau_n - \delta \tau / 2}^{\tau_n + \delta \tau / 2} \mathcal{C}_{I
\nocomma I}  (- i \tau_n) \mathd \tau$ over the time steps of our
discretization, which removes the weak divergence. This amounts to filtering
the correlation function by convolving it by a rectangular function, and hence
to multiply the Fourier coefficients $\mathcal{C}_n$ by a sinc
\[ \mathcal{C}_n \rightarrow \bar{\mathcal{C}}_n =\mathcal{C}_n \tmop{sinc}
   \frac{\delta \tau}{2} \omega_n =\mathcal{C}_n \tmop{sinc} \frac{n \pi}{M},
\]
and to define the $\{R_m \}$ from these $\{ \bar{\mathcal{C}}_m \}$.

Even with such filtering, a second problem remains : when taking the FFT, we
only sum the $M$ first Fourier coefficients so that the correlator we obtain
deviates from the ideal value, as apparent in Eq. \eqref{delta}. When $M$ is
large enough, this deviation leaves a noticeable systematic error only for the
same-time correlator
\begin{eqnarray*}
  \langle \xi (\tau_n) \xi (\tau_n) \rangle & = & \overline{\mathcal{C}_{I
  \nocomma I}} (0) - \Delta
\end{eqnarray*}
where the error $\Delta$ is
\[ \Delta = \sum^{\infty}_{j = M} \bar{\mathcal{C}}_j = \frac{\hbar}{R}
   \omega_c^2  \frac{M \tmop{Im} \left( \Phi \left( e^{- \frac{i \pi}{M}}, 1,
   M \right) - \Phi \left( e^{- \frac{i \pi}{M}}, 1, M + \frac{\beta \hbar
   \omega_c}{2 \pi} \right) \right)}{\pi \beta \hbar \omega_c} . \]
Such Dirac delta-like error can be easily corrected by applying a shift to all
the Fourier coefficients entering our FFT, except the zero-frequency one which
provides the correct baseline
\[ \bar{\mathcal{C}}_j \rightarrow \tilde{\mathcal{C}}_j = \bar{\mathcal{C}}_j
   - (1 - \delta_{0 j})  \frac{\Delta}{M - 1}, \qquad j = 0, \ldots, M - 1. \]
The $\{R_m \}$ are finally evaluated from the $\{ \tilde{\mathcal{C}}_m \}$ in
place of the initial $\{\mathcal{C}_m \}$. With these corrections made, we
compare the numerical correlations to the expected $\overline{\mathcal{C}_{I
\nocomma I}} (- i (\tau))$ in Fig. \ref{check_correl}

\section{Basis states, operator matrices and thermal expectation values for
the bare CPB}\label{CPBApp}

In this appendix we evaluate thermal expectations values of some operators of
the CPB, working in the eigenbasis. The CPB eigenstates can easily be obtained
numerically in a truncated discrete charge basis, but below we rather obtain
them analytically in terms of Mathieu functions
{\cite{cottet_implementation_2002-1,koch_charge_2007}}.

The Shrödinger differential equation for the bare CPB Hamiltonian in absence
of offset charge is
\begin{equation}
  E_C \Psi \text{''} (\varphi) - \left( E_J  \text{cos} \varphi \right) \Psi
  (\varphi) = E \Psi (\varphi) \label{EDCPB} .
\end{equation}
This equation is a form of Mathieu's equation
\[ f'' (z) + (a - 2 q \cos 2 z) f (z) = 0, \]
whose solutions are known as special functions {\cite{noauthor_nist_nodate}}.
Furthermore, given that the potential is periodic in $\varphi$, Bloch's
theorem implies the eigenfunctions of \eqref{EDCPB} are of the form
\[ \Psi_{n \nocomma p} (\varphi) = \langle \varphi |n, p \rangle = e^{ip
   \varphi} u_{n \nocomma p} (\varphi), \]
where $n$ is a band index, and $p$ is the quasicharge (i.e. Bloch's
quasimomentum), with $u_{n \nocomma p} (\varphi)$ a $2 \pi -$periodic function
of $\varphi$ (same period as the $\cos \varphi$ potential). If the CPB is not
connected to anything, $p$ is fixed to zero, whereas when connected to a
circuit that can let charge circulate, $p$ can fluctuate and take any value in
$\mathbb{R}$.

Using knowledge from the solutions of Mathieu's equation, the eigenenergy
$E_{n \nocomma p}$ of $\Psi_{n \nocomma p} (\varphi)$ is given by
\[ E_{n \nocomma p} = \frac{E_C}{4} \lambda_{\chi (n, p)}  (- 2 E_J / E_C), \]
where $\lambda_{\nu}$ denotes the \tmtextit{Mathieu characteristic value}
special function indexed by its \tmtextit{characteristic} \tmtextit{exponent}
and
\[ \chi (n, p) = n + n \bmod 2 + (- 1)^n 2 | \tmop{frac} (p) | \]
is a function giving the characteristic exponents, such that the eigenenergies
are sorted increasing with the band index $n \in \mathbb{N}$, and where the
fractional value $\tmop{frac} (p) = p - \tmop{round} (p)$, with $\tmop{round}
(p)$, rounding to the nearest integer. The $u_{n \nocomma p}$ functions
themselves can be expressed as
\[ u_{n \nocomma p} (\varphi) = \frac{e^{- ip \varphi}}{\sqrt{2 \pi}}  \left(
   \tmop{ce}_{\chi (n, p)} \left( \frac{\varphi}{2}, - 2 \frac{E_J}{E_C}
   \right) + i (- 1)^n \tmop{sign} (\tmop{frac} (p)) \tmop{se}_{\chi (n, p)}
   \left( \frac{\varphi}{2}, - 2 \frac{E_J}{E_C} \right) \right), \]
where the Mathieu $\tmop{ce}_{\nu}$ and $\tmop{se}_{\nu}$ are respectively
even and odd real functions of $\varphi$ {\cite{noauthor_nist_nodate}}. Note
that $\lambda_{\nu}$ has discontinuities when $\nu = \chi (n, p)$ is strictly
an integer (\tmtextit{i.e.} when $2 p$ is an integer), as well as either
$\tmop{ce}_v$ or $\tmop{se}_{\nu}$; the eigensolutions to consider at these
values in each band are then obtained as the limits when approaching the
discontinuity. Our expressions with Mathieu special functions extend those of
Ref. {\cite{cottet_implementation_2002-1}} to all quasicharges values, but
differ from those of Ref. {\cite{koch_charge_2007}}.

It is then straightforward to obtain the matrix elements of $N = - i
\frac{\partial}{\partial \varphi}$ and $N^2$,

\begin{align}
  \langle n, p | N | n', p' \rangle & = \delta (p - p')  \left( \delta_{n
  \nocomma n'} p - i \int_0^{2 \pi} d \varphi \hspace{0.17em} u_{n \nocomma
  p}^{\ast} (\varphi) \frac{du_{n' p}}{d \varphi} (\varphi) \right),
  \nonumber\\
  \langle n, p | N^2 | n', p' \rangle & = \delta (p - p')  \left( \delta_{n
  \nocomma n'} p^2 - 2 ip \int_0^{2 \pi} d \varphi \hspace{0.17em} u_{n
  \nocomma p}^{\ast} (\varphi) \frac{du_{n' p}}{d \varphi} (\varphi) \right.
  \nonumber\\
  & \phantom{= \delta (p - p') (\delta_{n \nocomma n'} p^2} \left. -
  \int_0^{2 \pi} d \varphi \hspace{0.17em} u_{n \nocomma p}^{\ast} (\varphi)
  \frac{d^2 u_{n' p}}{d \varphi^2} (\varphi) \right), \nonumber{} 
\end{align}

and those of any function $f (\varphi)$ are

\begin{align}
  \langle n, p | f (\varphi) | n', p' \rangle & = \delta (p - p')  \int_0^{2
  \pi} d \varphi f (\varphi)  \hspace{0.17em} u_{n \nocomma p}^{\ast}
  (\varphi) u_{n' p} (\varphi) . \nonumber{} 
\end{align}

\begin{figure}[h]
  \begin{center}
    {\hspace*{\fill}}
    
    \resizebox{13cm}{!}{\includegraphics{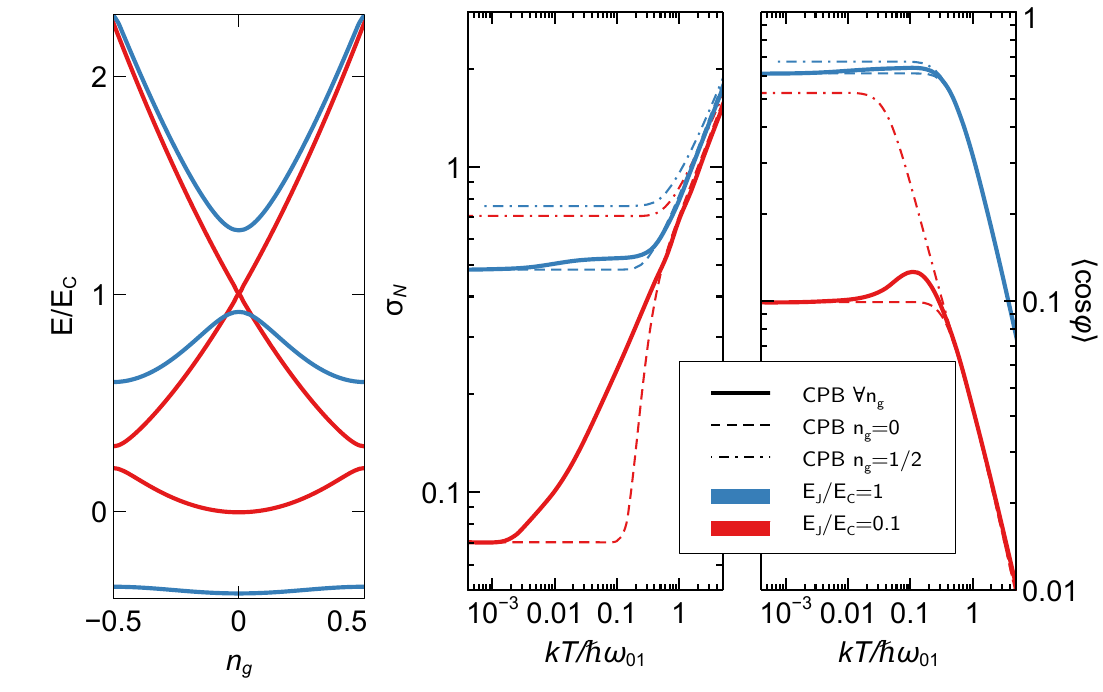}} 
  \end{center}
  \caption{\label{figCPB}For all panels, red: $E_J = 0.1 E_C$, blue: $E_J =
  E_C$ and $\hbar \omega_{01} = E_1  (n_g = 0) - E_0  (n_g = 0)$. Left panel:
  energy bands $E_0$, $E_1$ and $E_2$ (from bottom to top) of the CPB as a
  function of the gate charge. Middle (resp. right) panel, thermal expectation
  values of rms fluctuations of the charge $N$ (resp. Josephson coherence
  factor $\langle \cos \varphi \rangle$) as a function of temperature, for
  fixed gate charge $n_g = 0$ (thin dashed lines), $n_g = 1 / 2$ (thin
  dashed-dot lines), or (thick full lines) when allowing all gate charges. }
\end{figure}

Finally, we can evaluate thermal equilibrium expectation values from the
thermal density matrix $\rho_{\beta} = e^{- \beta H} / \tmop{Tre}^{- \beta H}$
and the matrix elements of operators as
\[ \langle A \rangle = \tmop{Tr} \rho_{\beta} A = \sum_n \int_{- 1 / 2}^{1 /
   2} dp \hspace{0.35em} e^{- \beta E_{np}} \langle n, p | A | n, p \rangle .
\]
In qubits, the quasicharge charge $p$ has values externally imposed by the
gate. The low impedance of the gate voltage is such that $p$ is nearly fixed
and one should then only sum over the band index. On the other hand, if the
qubit's ``island'' is not connected to a gate capacitance but rather to an
element that can let charge circulate, $n_g$ can fluctuate and take any value.
In Fig. \ref{figCPB}, assuming either fixed charge offset or that $p$ (or
$n_g$) can take any value, we plot the thermal expectation values $\sigma_N =
\langle N^2 \rangle^{1 / 2}$ of the rms fluctuations of the charge $N$, and
the Josephson coherence factor $\langle \cos \varphi \rangle$, which, being
non-zero, are both indicators of the superconducting character of the
unshunted CPB. One could also consider $\langle \sin^2 \varphi \rangle = (1 -
\langle \cos 2 \varphi \rangle) / 2$, the fluctuations of the supercurrent,
which is 1/2 when the junction is insulating ($\varphi$ is delocalized, with
all values equally probable) and smaller than 1/2 when the junction has finite
supercurrent fluctuations ($\langle \sin^2 \varphi \rangle = 0$ for the
classical superconducting junction in absence of phase bias).

This figure shows that at high temperatures, when $kT \gtrsim \hbar
\omega_{01}$ (temperature larger than the separation of the first two bands at
$n_g = 0$), the expectation values follow power laws $\sigma_N = \sqrt{kT / 2
E_C}$ and $\langle \cos \varphi \rangle \propto T^{- 1}$, with values
independent of whether $n_g$ is kept fixed or allowed to vary. In the opposite
low temperature limit where $kT \ll E_0  (n_g = 1 / 2) - E_0  (n_g = 0)$ (the
amplitude of the ground quasicharge band), expectation values saturate to a
plateau corresponding to the zero point fluctuations of the ground state at
$n_g = 0$. For $E_J \ll E_C$, these zero point values are $2 \sigma_N^2 =
\langle \cos \varphi \rangle \simeq E_J / E_C$.

In the intermediate temperature range, allowing charge fluctuations on the
capacitor enhances both $\sigma_N$ and $\langle \cos \varphi \rangle$ with
respect to the fixed $n_g = 0$ case, and this effect is most pronounced when
$E_C / E_J$ is large (deep ground quasicharge band). For $\langle \cos \varphi
\rangle$, this notably leads to a striking non-monotonic $T -$dependence, with
a local maximum at $kT \sim 0.1 \hbar \omega_{01}$. This maximum is easily
explained. In JJs with $E_C \gg E_J$, Cooper pair transfer occurs mostly
through a second order tunneling process of quasiparticles, with a virtual
intermediate state on higher charge parabolas. When allowing thermal
fluctuations of the quasicharge away from 0 in the ground band, the energy
difference between the ground and the lowest virtual excited state is reduced,
hence increasing the effective Josephson coupling. At temperatures $kT \sim
\hbar \omega_{01}$ or higher, the excited bands also get populated which then
reduces the effective Josephson coupling.

\section{Comparing our work to the literature}\label{Comparing}

\subsection{Effective actions}\label{actions}

The basis of our approach is the same as that used in most of the literature
on Schmid's transition. Starting from the Caldeira-Leggett Hamiltonian
\eqref{HCL} the dissipative bath is traced out using the Feynman-Vernon
influence functional to obtain an effective action, from which one infers the
properties of the system. However, several choices are possible, leading to
different writings for the effective action in different works. Here, we show
that we are describing the physics of the RSJ on the same grounds as in the
rest of the literature on Schmid's transition, although with a more general
kernel in the influence functional.

The total action for the system with the bath influence functional we
consider, before performing the Hubbard-Stratonovich transformation, is (Eq.
(\ref{SFl}-\ref{IF}))
\begin{equation}
  S_{\tmop{Fl}}^E [\varphi] + \Phi [\varphi] = \int_0^{\hbar \beta} d \tau
  (\frac{\hbar^2}{4 E_C}  \dot{\varphi}^2 - E_J \cos \varphi + E_L \varphi^2)
  \hspace{0.27em} - \frac{1}{2}  \int_0^{\hbar \beta} d \tau \int_0^{\hbar
  \beta} d \tau' \varphi (\tau) K (\tau - \tau') \varphi (\tau'),
  \label{ouraction}
\end{equation}
with the kernel $K$ given by Eqs. \eqref{Kernel} and \eqref{kernellorentz}. In
the review Ref. {\cite{schon_quantum_1990}}, Schön and Zaikin write the
effective action for the RSJ as
\begin{equation}
  \label{WTaction} S [\varphi] = \int_0^{\hbar \beta} d \tau (\frac{\hbar^2}{4
  E_C}  \dot{\varphi}^2 - E_J \cos \varphi) + \frac{1}{4}  \int_0^{\hbar
  \beta} d \tau \int_0^{\hbar \beta} d \tau' K_{\infty}  (\tau - \tau') 
  (\varphi (\tau') - \varphi (\tau))^2
\end{equation}
(converting their notations to ours) where the first integral is the action of
the Cooper pair box (with no counter-term) and where the kernel $K_{\infty}
(\tau)$ has the form
\begin{equation}
  K_{\infty} (\tau) = \frac{R_Q}{2 R}  \frac{\hbar}{\left( \hbar \beta \sin
  \frac{\pi \tau}{\hbar \beta} \right)^2} = \lim_{\omega_c \rightarrow \infty}
  K (\tau) \label{Knocutoff},
\end{equation}
which corresponds to the particular case where the kernel Eq. \eqref{Kernel}
is evaluated with a purely Ohmic admittance Re$Y (\omega) = 1 / R$, without
any UV cutoff ({\tmem{i.e.}} taking $\omega_c = \infty$ in \eqref{ReY}). The
action used by Schmid {\cite{schmid_diffusion_1983}} is the same as
\eqref{WTaction}, with the influence kernel being furthermore the zero
temperature limit of \eqref{Knocutoff}. For a moment, let us consider the
influence functional of \eqref{WTaction} with the more general kernel $K$ in
place of $K_{\infty}$, and expand the square of phase difference. Then, using
the facts that $K$ is even and periodic and that $\int_0^{\beta \hbar} d \tau
K (\tau) = 2 E_L$, one indeed formally recovers our form of the action with
the counter-term,
\begin{eqnarray}
  \frac{1}{4}  \int_0^{\beta \hbar} d \tau \int_0^{\beta \hbar} d \tau' K
  (\tau - \tau')  (\varphi (\tau') - \varphi (\tau))^2 & = & \int_0^{\beta
  \hbar} d \tau E_L \varphi (\tau)^2 \nonumber\\
  &  & - \frac{1}{2}  \int_0^{\hbar \beta} d \tau \int_0^{\hbar \beta} d
  \tau' \varphi (\tau) K (\tau - \tau') \varphi (\tau') . \label{equiv} 
\end{eqnarray}
Thus, provided one uses the same finite-cutoff kernel \eqref{Kernel} in our
effective action \eqref{ouraction} and in \eqref{WTaction}, the path integrals
are equal; in particular, the ground states of these actions coincide. Hence,
the localized ground state we find is also a valid ground state for Schmid's
action with the finite-cutoff kernel \eqref{Kernel}. This remains true when
taking limits (\tmtextit{e.g.} $T \rightarrow 0$), and in particular for the
$\omega_c \rightarrow \infty$ limit considered by many authors, where we find
this localized ground state becomes fully localized in phase (see Sec.
\ref{discussion}).

For completeness, we present another form of the action in Appendix
\ref{phas_invariance}. \

\subsection{Why our results contradict previous theoretical work}\label{why}

Even though our path integral equations appear compatible with those used by
previous authors confirming Schmid's QPT prediction (see \ref{actions}), our
equations lead us to conclude the opposite of these authors regarding that
QPT. We discuss here more specifically the work of Werner and Troyer (WT)
{\cite{werner_efficient_2005}}, who use the effective action \eqref{WTaction}
and apply the path integral quantum Monte Carlo numerical technique to
investigate the predicted QPT in the RSJ. Hence, in principle, their technique
and ours both calculate numerically the same path integral, and a close
examination should reveal why we come to different conclusions regarding the
predicted QPT.

First, we observe that in spite of the formal equivalence of our two methods,
it is not possible to directly recover and check WT's results with our
stochastic Liouville method because, with the infinite cutoff kernel
\eqref{Knocutoff} they use, (i) $E_L = \frac{1}{2}  \int_0^{\beta \hbar} d
\tau K (\tau) = \infty$ and (ii) whatever the time discretization chosen, the
strong $\tau^{- 2}$ divergence of $K$ at short times prevents drawing small
stochastic increments for a proper numerical integration of the Liouville
equation. This unexpected non-equivalence of our two methods in the infinite
cutoff limit considered by WT highlights that this limit requires careful
handling (something one can hardly do by considering from the onset an
infinite cutoff).

Precisely, in Sec. \ref{sec:results} and \ref{discussion} of the main text, we
investigate the role of the bath cutoff $\omega_c$ and show that, for reasons
easily explained, phase fluctuations reduce as $\omega_c$ increases,
eventually reaching a classical phase state in the $\omega_c \rightarrow
\infty$ limit. This trend and this limit we find are the opposite of what
would be needed to recover WT's numerical results when $R > R_Q$,
{\tmem{viz.}} a state with diverging phase fluctuations (see Fig. 3 in
{\cite{werner_cluster_2004}}, the preprint version of
{\cite{werner_efficient_2005}}). We also find a result opposite to them when
we evaluate the linear response as they do (see next subsection \ref{Kubo},
and Fig. 3 of Ref. {\cite{werner_efficient_2005}}). We stress that, on our
side, we take the limit in a controlled manner and that our results are in
agreement with Kubo's linear response, with symmetry considerations, with
experiments, and free of the issues mentioned in Appendix \ref{history}. All
the contrary for WT's results which only agree with Schmid's results.

However, we observe that these opposite results are obtained by taking limits
differently : in our approach, we first take the low $T$ limit and then we can
consider the infinite cutoff limit, while WT take the same limits in the
reverse order. The different outcomes reveal that these two limits do not
commute. Highlighting this non-commutation of limits is a key result of our
work which enables resolving the conundrum around Schmid's prediction.

In summary, by choosing to use the kernel \eqref{Knocutoff} in the action
\eqref{WTaction}, one implicitly assumes (i) an infinite cutoff would
correctly describe an actual RSJ experiment, and also supposes (ii) the
environment cutoff can be taken to infinity before evaluating the path
integral and considering its low temperature limit. Our work reveals that
neither of these implicit assumptions holds. These subtle unfulfilled
assumptions suffice to explain why the phase delocalization QPT found by WT,
(confirming Schmid's results) does not describe the actual physics: it is an
uncontrolled, non-physical result in that model of the RSJ. These unfulfilled
assumptions impact similarly the results of all the other authors (Schmid, in
particular) who consider the same $\omega_c = \infty$ limit from the onset.

It happens that the (nonphysical) results obtained in this model by first
taking the non-commuting $\omega_c = \infty$ limit {\tmem{coincide}} with the
QPT phenomenology of the 1D quantum impurity problem
{\cite{kane_transport_1992,torre_viewpoint_2018}}. Before the present work, it
was claimed that these systems were equivalent and this coincidence of results
was seen as a strong validation of Schmid's prediction in this model (and the
whole literature on it). Below, in \ref{kane-fisher}, we explain why the 1D
systems and the Caldeira-Leggett model of the RSJ are in fact {\tmem{not}}
equivalent (one indeed having a QPT, the other, not).

\subsection{Evaluating dc mobility using the stochastic Liouville
method}\label{Kubo}

\begin{figure}[h]
  \resizebox{10cm}{!}{\includegraphics{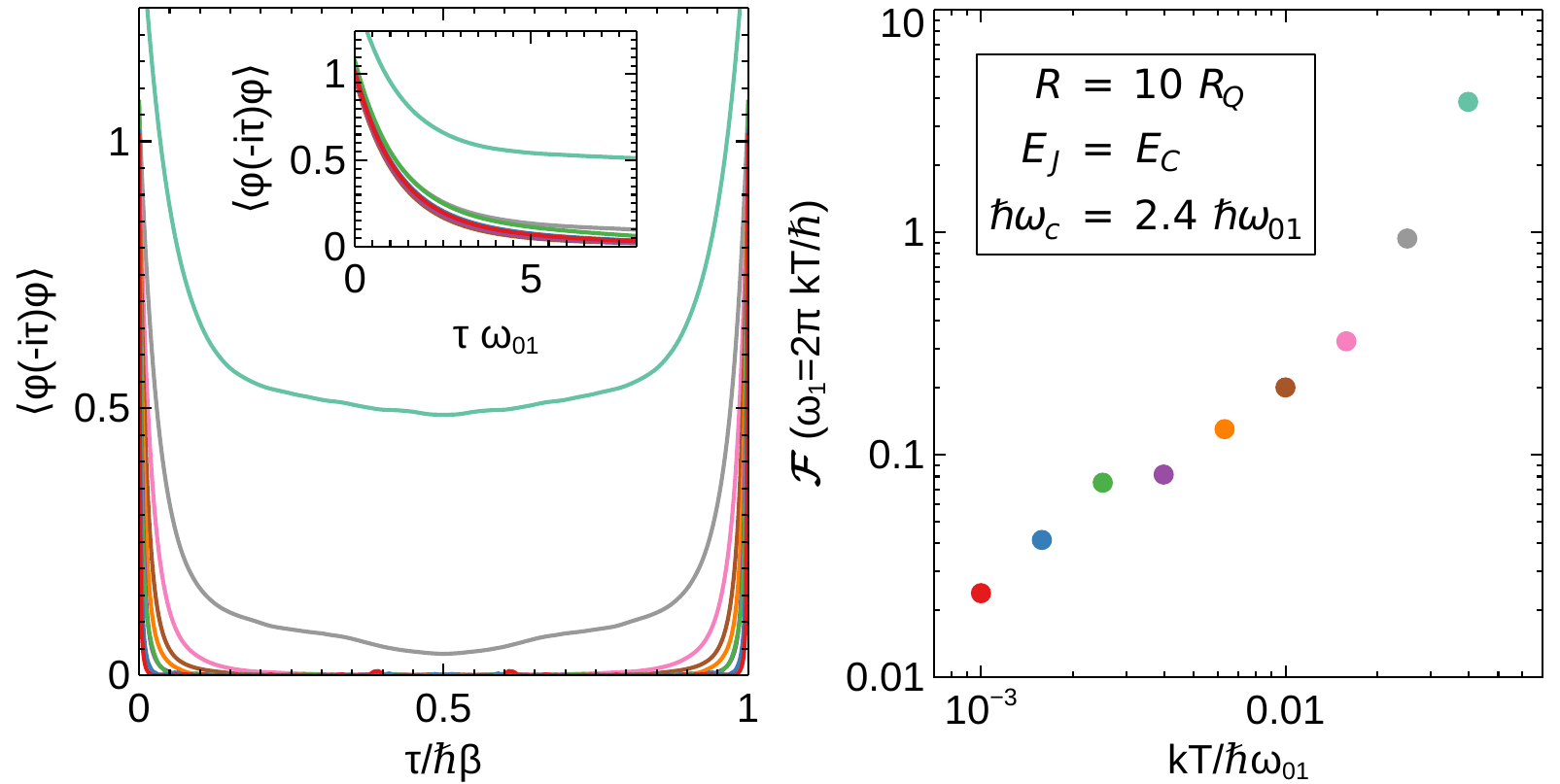}}
  \caption{\label{correl_mob}Imaginary time dynamics of the system from
  stochastic Liouville simulations, for the same RSJ parameters as in Fig.
  \ref{lin_resp}. Left panel : imaginary time correlator of the phase, for
  different temperatures (the temperature of a given curve is the abscissa of
  the same-colored dot in the right panel). Main panel : full period of the
  correlator. Inset : short-time close-up with a fixed timescale, showing that
  the decay of the correlator becomes $T$-independent at low temperatures, as
  the system reaches its ground state. Right panel : evolution of Eq.
  \eqref{limitWT}, calculated with the left panel data. The saturation shown
  in the left panel inset, makes the integral in \eqref{limitWT} tend to a
  constant, and thus, up to the uncertainty due to the stochastic method, this
  quantity varies linearly with the first Matsubara frequency $\omega_1$ at
  low temperature, clearly pointing to $\tmop{Re} Z (0) = 0$ for the RSJ. When
  increasing $\omega_c$, $\langle \varphi \varphi \rangle$ is reduced and this
  result persists. \ \ \ }
\end{figure}

In Ref. {\cite{werner_efficient_2005}}, WT take advantage of the analyticity
of $S_{\varphi \varphi}  (\omega)$ to evaluate dc impedance of the RSJ as
$\lim_{\omega \rightarrow 0} (\omega S_{\varphi \varphi}  (\omega))$ (see
\ref{KuboTh}) by extrapolating the variations of

\begin{equation}
  \mathcal{F}  (\omega_1) = \omega_1 S_{\varphi \varphi}  (i \omega_1) =
  \frac{\omega_1}{\hbar} \int_0^{\hbar \beta} \langle \varphi (- i \tau)
  \varphi \rangle e^{i \omega_1 \tau} d \tau \label{limitWT}
\end{equation}

upon reducing the first Matsubara frequency $\omega_1 = 2 \pi k T / \hbar$
{\cite{werner_efficient_2005}}. The stochastic Liouville method can also
provide the imaginary time correlator $\langle \varphi (- i \tau_m) \varphi
\rangle$ at the discrete time steps $\{ \tau_m = m \delta \tau \}$ (see Sec.
\ref{implementation}). For a given realization of the random noise $\xi
(\tau)$ and a given intermediate time $\tau_m, \tmop{Tr} \rho_{\xi} \varphi (-
i \tau_m) \varphi$ is obtained by splitting the discrete integration of the
imaginary time evolution in two parts, and inserting the phase operator both
at $\tau_0 = 0$ and at the split $\tau_m$, and taking the trace. Finally, as
for the RDM, for each intermediate time $\tau_m$, one averages over the values
obtained for the different realizations of $\xi$. In Fig. \ref{correl_mob}, we
carry out the same extrapolation as WT on our numerical results for the
imaginary time correlator $\langle \varphi (- i \tau) \varphi \rangle$, for $R
= 10 R_Q$. The data clearly point to a vanishing limit for \eqref{limitWT},
\tmtextit{i.e.} a superconducting linear response, consistent with the
rigorous time-domain result found in \ref{KuboTh} and all our other results.
This contradicts the results in Fig. 3 of WT {\cite{werner_efficient_2005}}
for $R > R_Q$ (see discussion above in \ref{why})

\subsection{Why the RSJ differs from 1D quantum impurity
problems}\label{kane-fisher}

In a famous paper {\cite{kane_transport_1992}}, Kane and Fisher (KF) showed
that a 1D electronic channel with an impurity has a QPT dependent on the
Luttinger interaction parameter in the channel, and the scaling laws of the
corresponding quantum critical regime were indeed observed experimentally
{\cite{jezouin_tomonaga-luttinger_2013,parmentier_strong_2011,anthore_circuit_2018}}
in related systems. In their work, by bosonizing the 1D fermions
{\cite{von_Delft_bosonization_1998,schulz_fermi_2000}}, KF derive equations
they present as ``formally equivalent to the Caldeira-Leggett model for a
resistively shunted Josephson junction'', with notably a cosine term similar
to the Josephson coupling, and they identify their QPT with Schmid's QPT. The
fact that our present theoretical results exclude a QPT for the RSJ in this
model (in agreement with experiment), while the QPT exists both in the theory
and experiments on the 1D impurity problem, shows that, actually, the
{\tmem{equivalence}} does not hold. Here, we explain why the cosine term they
obtain is, in fact, not analogous to the Josephson coupling; it describes a
different physics.

The cosine term obtained by KF describes the scattering, by the impurity, of
bosonic excitations which act as the dissipative bath in the 1D systems. These
excitations are defined only for a non-zero wavevector
{\cite{von_Delft_bosonization_1998,schulz_fermi_2000}}, and thus have a
strictly positive energy. Hence, this cosine term has an implicit IR cutoff;
notably its matrix element in the ground state of these 1D systems is zero.

On the other hand, in the Caldeira-Leggett Hamiltonian \eqref{HCL} we use to
model the RSJ, the $- E_J \cos \varphi$ term is an effective potential due to
coherent tunneling between the BCS ground states on both sides of the
junction; as such, it has no IR cutoff. Its argument, the phase $\varphi$, is
a degree of freedom of the Hamiltonian \eqref{HCL} and, in this model, its
amplitude $E_J$ is considered independent of the dissipative admittance (in
Schmid's paper {\cite{schmid_diffusion_1983}} too, the potential depth is
independent of the dissipation) and given by the Ambegaokar-Baratoff value
{\cite{ambegaokar_tunneling_1963}}. To be more specific, this term in the
Hamiltonian exists with or without a dissipative bath (e.g. in the bare CPB
with the Hamiltonian ${H_{\tmop{CPB}}} $ \eqref{HCPB}) and it always confers
to the junction its inductive (see \eqref{eomidot} and Sec. \ref{KuboTh}),
superconducting, behavior. In the presence of the ohmic bath, this $- E_J \cos
\varphi$ potential entails the infinitely degenerate ground state (see App.
\ref{symmetries}). In contrast, in the 1D systems, KF's cosine term cannot be
dissociated from the bath. Its argument is a sum of bath variables which is
irrelevant in the ground state; it is thus not an independent degree of
freedom characterizing the state of the system, unlike $\varphi$ in
\eqref{HCL}. The 1D impurity problem therefore lacks the RSJ ground state
properties that ensue from the existence of this degree of freedom (e.g. the
inductive response).

Even though we can explain why 1D systems behave differently from the RSJ in
spite of deceptively similar-looking equations, it is not clear to us why
considering the (non-commuting) $\omega_c = \infty$ limit from the onset in
the RSJ (as Schmid, WT, and most authors did) yields in practice the QPT
phenomenology of the 1D systems (see \ref{why}). How come taking this
non-commuting limit first (and thus, without control on the results) amounts
to unwittingly suppress the degree of freedom $\varphi$? At this point, it
seems just an unfortunate coincidence.

\section{Symmetries and degeneracies in the RSJ}\label{symmetries}

\subsection{Phase translation invariance}\label{phas_invariance}

The Caldeira-Leggett Hamiltonian \eqref{HCL} is left invariant by the discrete
translation symmetry that simultaneously shifts the junction's phase and all
the bath oscillators' phases by the same multiple of $2 \pi$:
\[ \left. \begin{array}{rcl}
     \varphi & \rightarrow & \varphi + 2 \pi k\\
     \forall n, \quad \varphi_n & \rightarrow & \varphi_n + 2 \pi k
   \end{array} \right\} k \in \mathbb{Z} \]
After tracing out the bath oscillators, a translational invariance with
respect to the sole junction phase
\begin{equation}
  \varphi \rightarrow \varphi + 2 \pi k, k \in \mathbb{Z} \label{DTI}
\end{equation}
must remain for the states of the RSJ obtained from the RDM. Indeed, using
\eqref{WTaction} and \eqref{equiv} one can check this invariance is present in
our path integral \eqref{RDMaction}, before the Hubbard-Stratonovich
transformation is made. However, the way the Hubbard-Stratonovich
transformation is implemented, replacing a quadratic term in $\varphi$ by a
linear term and leading to Eq. (\ref{PI2}-\ref{Sfict}), breaks the above
translation invariance. As a result, Eq. (\ref{PI2}-\ref{Sfict}) implicitly
select states that are centered at $\varphi = 0$. The $k \neq 0$ translated
states would be obtained as solutions of the translated version of Eq.
(\ref{PI2}-\ref{Sfict}), where the potential terms in the fictitious system
are translated. The effective action we use in the main text thus only yields
a subset of the system states, and these localized states are infinitely
degenerate through the $k$ values of this translation.

Although it may seem incorrect at first sight that our localized solutions do
not have the translational symmetry builtin the original equations, this is
neither an error nor a problem. Indeed, it is well known that when the
eigenstates of a system are degenerate, some eigenstates may have a lower
symmetry than the system as a whole (think \tmtextit{e.g.} of the $\ell > 0$
orbitals of the hydrogen atom, which do not have the spherical symmetry of the
atom's Hamiltonian). A complete basis of the degenerate subspace has the
appropriate symmetries.

Let us consider the ground state RDM $\rho_0 = | \Psi_0  \rangle \langle
\Psi_0  |$ obtained from Eq. (\ref{PI2}-\ref{Sfict}), defining a reduced
ground state $| \Psi_0  \rangle$ localized and centered at $\varphi = 0$ (the
diagonal $\langle \varphi | \rho_0 | \varphi \rangle$ of this RDM is the
square modulus of the reduced wave function $\Psi_0 (\varphi) = \langle
\varphi | \Psi_0  \rangle$, \tmtextit{i.e.} the density of probability of the
phase in this ground state). Then, following the above translation arguments,
all the translated states  \{$e^{- i 2 \pi k \hat{N}} | \Psi_0  \rangle, k \in
\mathbb{Z}$\}  are also valid reduced ground states of the RSJ. From the
ensemble of these translated states, one can furthermore construct fully
delocalized Bloch states with a dimensionless quasicharge $q$
\begin{equation}
  | \Phi_q  \rangle =\mathcal{N} \sum_k e^{- i 2 \pi k (q + \hat{N})} | \Psi_0
  \rangle \label{Bloch}
\end{equation}
(with $\mathcal{N}$ the usual normalization factor of Bloch states) which have
the same energy as $| \Psi_0  \rangle$ for all values of $q$, and are hence
also degenerate ground states. Thus, unlike the CPB, the RSJ has a flat
quasicharge ground band, as already stated in Refs.
{\cite{murani_reply_2021,murani_absence_2020}} (see also next subsection
\ref{CTI}). The fact that we can exhibit both localized and delocalized ground
states illustrates the abstract argument given in Ref.
{\cite{murani_absence_2020}}, that any valid equilibrium state of the RSJ
ought to be representable both as a localized or a delocalized state, based on
formal arguments on unitary transformations developed earlier in Ref.
{\cite{loss_effect_1991-1,mullen_resonant_1993}}.

\begin{figure}[h]
  \resizebox{12cm}{!}{\includegraphics{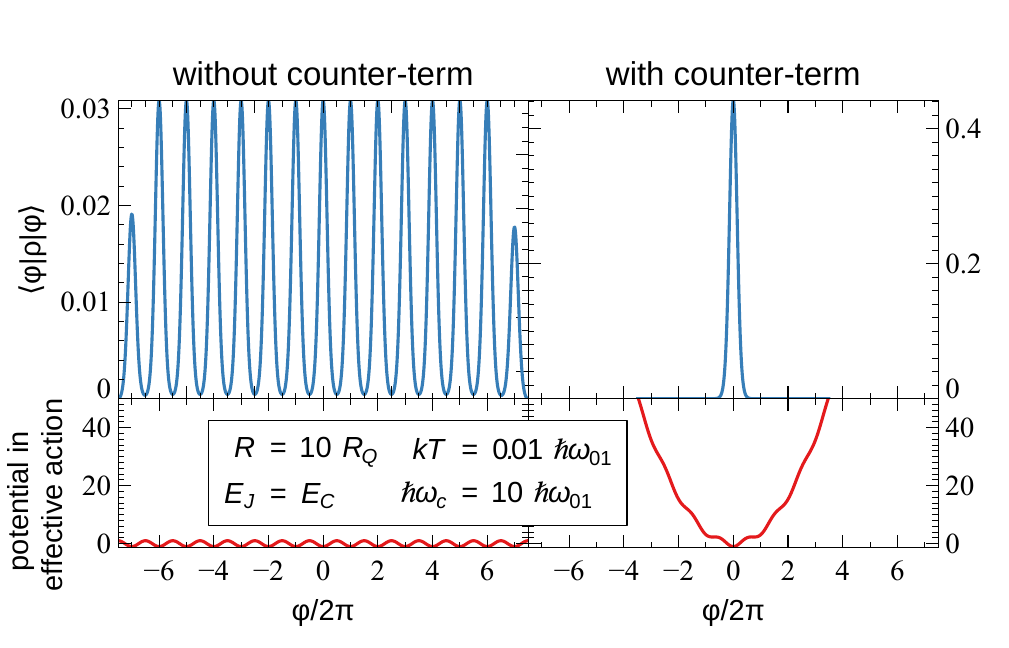}}
  \caption{\label{ct_vs_noct}For the same set of parameters given in the
  label, we compare of the probability density $\langle \varphi | \rho |
  \varphi \rangle$ where $\rho$ is the RDM obtained from the stochastic
  Liouville method, computed with either (right panels) the effective action
  with the counter-term \eqref{ouraction} used in this paper, or (left panels)
  the modified effective action \eqref{altaction} without the counter-term.
  The counter-term-free solution is close to a Bloch-like equal-weight
  superposition of repeated $2 \pi$-shifted copies of the localized solution
  obtained with the counter-term, up to edge effects due to the limited number
  of cosine wells considered in the computation. }
\end{figure}

This localized\textbar delocalized duality is further confirmed by the fact
that one can come up with different (yet equivalent) equations for that system
and which directly yield only delocalized states, for all parameters. Notably,
it is possible to transform our effective action \eqref{ouraction}
($\Leftrightarrow$ Eq. (\ref{SFl}-\ref{IF})) to a mathematically equivalent
form :
\begin{equation}
  S_{\tmop{CPB}}^E [\varphi] + \tilde{\Phi} [\varphi] = \int_0^{\hbar \beta} d
  \tau (\frac{\hbar^2}{4 E_C}  \dot{\varphi}^2 - E_J \cos \varphi)
  \hspace{0.27em} - \frac{1}{2}  \int_0^{\hbar \beta} d \tau \int_0^{\hbar
  \beta} d \tau' \varphi (\tau) k (\tau - \tau') \varphi (\tau'),
  \label{altaction}
\end{equation}
where the parabolic counter-term potential in the first term has been removed
(\tmtextit{i.e.} it is the action of a CPB instead of a fluxonium) and
absorbed in the modified influence functional $\tilde{\Phi}$ (See e.g.
{\cite{weiss_quantum_2012}} or {\cite{grabert_quantum_1988}}). The kernel $k$
of this modified influence functional is related to the kernel $K$
\eqref{K}-\eqref{Kernel} of the influence functional we use in the rest of
this work through \
\[ k (\tau) = K (\tau) - 2 E_L  \Sha_{\hbar \beta} (\tau), \]
where the $ \Sha_{\hbar \beta} (\tau) = \sum_{n = - \infty}^{+ \infty} \delta
(\tau - n \hbar \beta)$ is the Dirac comb of period $\hbar \beta$, such that
$\int_0^{\beta \hbar} d \tau k (\tau) = 0$. This alternate writing of the
effective action does not have the confining potential that localizes our
states, and that may seem artificial to some. After performing the
Hubbard-Stratonovich transformation on this effective action, the fictitious
system is a particle in a periodic potential submitted to a random noise which
causes the particle to diffuse. In equilibrium, the diffusion current must
vanish and the probability to find the particle is equal in all the wells.
This is confirmed by running the stochastic Liouville method (see Sec.
\ref{SLEsec}) with this counter-term-free modified action, as shown in Fig.
\ref{ct_vs_noct}. We observe that, indeed, without the counter-term, the
results resemble the delocalized Bloch states described above, with
expectations values of observables matching those of the simulations performed
with the counter-term, within the error bars of the simulations. Were the
number of wells taken into account in the numerics to be increased, the ground
state obtained with this modified kernel would tend to the perfectly periodic
phase state $| \Phi_{q = 0}  \rangle$ (also known as a compact phase state),
as reasoned above.

An important additional consequence of the translational invariance
\eqref{DTI} of the effective action is that quantities that are actually
measurable on that system can only involve $2 \pi$-periodic functions of
$\varphi$. Thus, the phase itself, and all its moments, are {\tmem{not}}
measurable.

\subsection{Charge translation invariance}\label{CTI}

A unitary transformation $U = \exp \left( i \varphi \sum_n N_n \right)$
applied to the Caldeira-Leggett Hamiltonian \eqref{HCL} yields the so-called
charge gauge Hamiltonian $E_C \left( N - \sum_n N_n \right)^2 - E_J \cos
\varphi + H_{\tmop{bath}}$. The later Hamiltonian is invariant upon a
translation of $N$ and any of the bath $N_n$ by the same arbitrary amount $\in
\mathbb{R}$. Similarly to the phase translations, this means that, after
tracing out the bath, for the reduced ground state $| \Psi \rangle $obtained
from the ground state RDM, any charge-translated copy $e^{i q \varphi} | \Psi
\rangle$, $q \in \mathbb{R}$ is also a valid reduced ground state, with the
same observables. This is another way to establish the degeneracy of the
ground states with respect to the quasicharge, i.e. the flatness of the
quasicharge ground band. In other words, one can circulate an arbitrary charge
in the loop formed by the junction and the resistance, without changing the
properties of the system's ground state.

\subsection{In the RSJ, ``the'' ground state has no definite symmetry with
respect to the phase.}

The material in this Appendix shows that the infinite ground state degeneracy
of the RSJ makes it pointless to debate about what ought to be ``the''
symmetry of ``the'' ground state with respect to the junction phase. This
should settle the longstanding debate on wether a galvanically-connected
non-superconducting environment forces one to consider the JJ phase as an
``extended'' variable (as opposed to ``compact'' in the CPB). As explained
above, the junction phase in the RSJ is not measurable; there is no actual
physical meaning associated to the localized or delocalized symmetry of a
given state in the phase representation, and, correspondingly, no measurement
can assess whether it is localized or delocalized, extended or compact. It can
also be regarded as the reason why Schmid's prediction of a
localization\textbar delocalization transition was erroneous: in this system,
it does not make much sense. The translation invariance and the ensuing
multiplicity of the ground states in the RSJ clearly differentiate this system
from the akin spin-boson problem regarding the possibility of a spontaneous
symmetry breaking caused by dissipation.

\subsection{Recovery of the CPB physics in the $R \rightarrow \infty$
limit}\label{recovery}

The CPB has a quasicharge ground band with a finite depth $E_S > 0$ (the
so-called phase slip energy) and a 2e-periodicity which reflects the charge
quantization on the island of the device. At a non-integer quasicharge $q$ on
this band, there is a finite constant charge on the capacitor (and there is a
voltage across it). In this state of the device, if we now connect a resistor
across the junction, the resistor drains away the charge on the capacitor and
the newly made RSJ eventually reaches its flat-band ground state at times
scales longer than $R C$. However, in the $R \rightarrow \infty$ limit this
relaxation takes an infinite time and one permanently observes the CPB
physics, with a static gate charge.

In other words, static charging effects (often called Coulomb blockade
effects) only occur in systems with an island. In the RSJ, a resistor with $R
< \infty$ suppresses the CBP island, making the ground band flat. But even
when the circuit no longer has an island, dynamical charging effects may still
be observed in it (\tmtextit{e.g.} by driving it non-adiabatically in dc or
ac, see \ref{stability}), similar to dynamical Coulomb blockade in island-less
normal-state circuits. Note however that, in spite of having a genuine island,
the CPB is dc-superconducting for a small enough current (see \ref{KuboTh}),
even though the curvature of its quasicharge ground band could suggest to
regard it as a capacitor.

Could it be that for $R_Q < R < \infty$, the symmetries of the Hamiltonian get
spontaneously broken, such that the ground state is no longer invariant w.r.t.
to charge translation, resulting in a non-flat quasicharge ground band? As
just discussed, such non-flat ground band indicates the system has an island,
and that would imply that the resistor became insulating, which would
contradict the behavior assumed in the first place for the resistor.

\subsection{Stability of the superconducting ground state at finite current
bias}\label{stability}

In this subsection, we critically examine the superconducting response of the
RSJ to a dc current bias, obtained in Sec. \ref{sec:results} (Fig.
\ref{lin_resp}) using stochastic Liouville, and discussed in Sec.
\ref{discussion}. The Hamiltonian of the dc-current-biased RSJ is $H = H_0 -
\varphi_0 I_b \varphi$. At $I_b \neq 0$, the potential of this Hamiltonian is
not bounded from below. In that case, the localized states we obtain from our
stochastic Liouville method, slightly off-centered from $\varphi = 0$, can, at
best, only be metastable (just as in the standard tilted washboard image of
current-biased JJs). However, in our approach, no runaway to a dissipative
state can occur because the states are confined by the counter-term. How
confident can we be that this superconducting linear response result we find
is not merely an artifact of the method?

By applying a time-dependent unitary transformation $U (t) = e^{- i I_b
\varphi t / 2 e}$ to the above current-biased Hamiltonian, we obtain the
transformed time-dependent Hamiltonian
\[ \tilde{H} = UHU^{\dagger} + i \hbar \dot{U} U^{\dagger} = E_C  \left( N +
   \frac{I_b}{2 e} t \right)^2 - E_J \cos \varphi + \sum_n 4 e^2 
   \frac{N_n^2}{2 C_n} + \varphi_0^2  \frac{(\varphi_n - \varphi)^2}{2 L_n} \]
where the potential is bounded as in \eqref{HCL} and where the bias current
now appears as a linear-in-time ``offset charge'' for the CPB. The above Bloch
state $| \Phi_q  \rangle$ \eqref{Bloch} with the quasicharge $q = I_b t / 2 e$
is an exact instantaneous (reduced) ground state for $\tilde{H}$ at time $t$.
Hence, as long as it can follow adiabatically its flat quasicharge ground
band, the system will remain in a zero-voltage state with the junction
sustaining the (super)current flow. Such adiabaticity is guaranteed by general
theorems {\cite{avron_adiabatic_1999,burgarth_one_2022}} at vanishing
$\dot{q}$, and the superconducting linear response independently worked out in
\ref{KuboTh} confirms that. The linear-response-displaced localized states
obtained with our stochastic Liouville method in Fig. \ref{lin_resp}
correspond to this adiabatic evolution; they are valid metastable solutions,
at least at vanishing bias. This confirms that the dc $I - V$ characteristic
of the RSJ at equilibrium is vertical at the origin.

However, when increasing the current bias $I_b$ (or $\dot{q}$), at some point,
adiabatic evolution in the ground band is no longer possible (in any case, one
can expect a hard limit at $| I_b | = I_0$); the metastable localized ground
state then occasionally experiences phase slips due to macroscopic quantum
tunneling through the cosine potential barrier or thermal activation over it,
and the junction state gradually or suddenly departs from zero voltage,
depending on whether the phase motion is over- or under-damped by the
resistance. This physics of the supercurrent branch was broadly understood
{\cite{ivanchenko_josephson_1969,mccumber_effect_1968,stewart_current-voltage_1968}}
well before Schmid's work. Caldeira and Leggett completed the understanding of
this loss of adiabaticity by evaluating quantitatively the phase slip rate out
of the metastable states for all parameters {\cite{caldeira_quantum_1983}}.
They showed that dissipation reduces quantum tunneling of the phase, thereby
increasing the lifetime of the metastable state and the ability of the
junction to sustain a supercurrent. This is consistent with our conclusion in
Sec. \ref{discussion} that, at equilibrium, a junction shunted by a resistor
is always more superconducting than an unshunted junction.

If one now considers an ac charge drive with a fixed finite amplitude but
variable frequency, the above superconducting adiabatic dynamics can only
occur at low enough frequency. In that regime, the junction behaves
inductively, with an inductance given by the linear response value
$\eqref{Leff}$. At sufficiently high frequency adiabaticity is lost, and the
behavior is the capacitive response of the CPB ground band at zero offset
charge. In-between, one expects a crossover at a given frequency (depending
non-linearly on the drive amplitude) with, at that point, likely, the response
of the resistor. This explains why uncontrolled noise in experiments may spoil
the thermal equilibrium superconducting linear response and the observation of
a supercurrent branch at the origin. Similar crossovers from inductive to
capacitive can be expected when varying other parameters.

\tmdetailed{bib style: SciPost\_bibstyle or unsrturl or tm-abstract}{\ }


\begin{thebibliography}{10}
\bibitem[43]{note-Gaussian-results}This Gaussian result (where the $2 \pi$
  translation invariance of the phase is lost) suggests that the effective
  Josephson coupling can be strongly suppressed by phase fluctuations
  occurring across large impedances. Our present exact results show that,
  actually, the effective josephson coupling $E_J \langle \cos \varphi
  \rangle$ has the lower bound $E_J^2 / E_C$.
\end{thebibliography}
\end{document}